\documentclass[preprint2]{aastex}

\usepackage{xcolor} 

\usepackage[colorlinks=true,breaklinks=true,linkcolor=magenta,citecolor=blue,urlcolor=green]{hyperref}
\newcommand\ergs{erg~s$^{-1}$}

\begin{document}

\shorttitle{Precession of supercritical accretion flow}
\shortauthors{Weng \& Feng}
\title{Evidence for precession due to supercritical accretion in ultraluminous X-ray sources}

\author{
Shan-Shan Weng\altaffilmark{1},
Hua Feng\altaffilmark{2}}

\altaffiltext{1}{Department of Physics and Institute of Theoretical Physics, Nanjing Normal University, Nanjing 210023, China}
\altaffiltext{2}{Department of Engineering Physics and Center for Astrophysics, Tsinghua University, Beijing 100084, China}

\begin{abstract}

Most ultraluminous X-ray sources (ULXs) are thought to be supercritical accreting compact objects, where massive outflows are inevitable. Using the long-term monitoring data with the {\it Swift} X-ray Telescope, we identified a common feature in bright, hard ULXs: they display a quasi-periodic modulation in their hard X-ray band but not in their soft band. As a result, some sources show a bimodal distribution on the hardness intensity map. We argue that these model-independent results can be well interpreted in a picture that involves supercritical accretion with precession, where the hard X-ray emission from the central funnel is more or less beamed, while the soft X-rays may arise from the photosphere of the massive outflow and be nearly isotropic. It implies that precession may be ubiquitous in supercritical systems, such as the Galactic microquasar SS~433.  How the hard X-rays are modulated can be used to constrain the angular distribution of the hard X-ray emission and the geometry of the accretion flow.  We also find that two ULX pulsars (NGC 5907 ULX-1 and NGC 7793 P13) show similar behaviors but no bimodal distribution, which may imply that they have a different beaming shape or mechanism.

\end{abstract}


\section{Introduction}

Ultraluminous X-ray sources (ULXs) are non-nuclear X-ray objects found in nearby galaxies \citep[for reviews, see][]{Feng2011,Kaaret2017}. Their apparent luminosities exceed the Eddington limit of a 20 $M_\odot$ black hole, or $3 \times 10^{39}$~\ergs, suggesting that they may contain black holes more massive than those found in Galactic X-ray binaries \citep{Colbert1999}. However, the majority of ULXs may be super-Eddington accreting objects, resting on several observational evidence. First, three ULXs were identified to contain a  neutron star accretor via the detection of coherent pulsations, while their apparent luminosities are well above the Eddington limit of a neutron star \citep{Bachetti2014, Fuerst2016, Israel2017,Israel2017a}. Second, on the perspective of X-ray spectra and spectral evolution, most ULXs do not behave in a manner similar to Galactic X-ray binaries at their sub-Eddington states \citep{Feng2009,Gladstone2009,Kajava2009,Bachetti2013,Sutton2013,Middleton2014,Walton2014}. Plus, direct evidence for fast outflows \citep{Pinto2016,Walton2016a}, along with the fact that some of the surrounding nebulae are due to energetic outflows interacting with the environment \citep{Pakull2003}, argues in favor of the supercritical accretion scenario \citep{Poutanen2007,Ohsuga2011,Jiang2014,Scadowski2016}. Thus, ULXs offer good opportunities for studying the accretion physics at extreme conditions.

As the accretion rate exceeds the Eddington limit, an analytical solution for the  accretion flow becomes increasingly difficult \citep{Abramowicz1978,Abramowicz1988,Begelman2002,Gu2016}, but analytical analysis indicates that the disk must be inflated and powerful winds will be launched due to radiation pressure \citep{Shakura1973,King2003,Poutanen2007,Shen2015}. Numerical simulations generally confirm the presence of massive outflows despite inconsistent details between different groups \citep[e.g.,][]{Ohsuga2011,Jiang2014,Narayan2017}.

Such a strong radiation pressure and outflow may have the natural consequence that it  may have a significant impact back on the accretion disk and lead to precession. A remarkable example is the Galactic microquasar SS~433, which, though not a ULX, though, undergoes an extremely high accretion rate ($\sim$$10^3$ times the Eddington limit) and shows strong outflows and obvious jet precession \citep{Fabrika2004}.  A possible analog of ULX may be M81 ULS-1, whose jet displayed a varying line-of-sight speed, and precession is a likely explanation \citep{Liu2015}.

As most ULXs do not show jets with atomic lines allowing for precession  measurement,  we thereby try to detect the signature of precession via its impact on the flux and spectral modulation.  So far, {\it Swift} has monitored about a dozen of the ULXs covering a timescale of up to several years, leaving behind a legacy needed for such searches. In this paper, we collected monitoring data for nine ULXs with the X-ray Telescope (XRT) on {\it Swift}. The data analysis and results are presented in Section~\ref{sec:data}, and the physical implications are discussed in Section~\ref{sec:dis}. We will show that, via a model-independent analysis, disk precession may be ubiquitous in ULXs.

\section{Data and results}
\label{sec:data}

The {\it Swift} satellite has been scheduled to monitor bright ULXs, accumulated with thousands of observations. Most of these monitoring programs are dedicated for the study of long-term variability of ULXs including the search of periodic or quasi-periodic signals \citep[e.g.,][]{Kaaret2009,Strohmayer2009,Grise2013,Lin2015,Walton2016,Hu2017}. For a single observation, the typical exposure time is 1--3~ks, and the typical recurrent time between observations is 2--6~days.  We collected ULXs that have been observed with {\it Swift} for at least 200 times as of 2017 August, but excluded those in M82 because they cannot be spatially resolved with XRT.  We also included the two known ULX pulsars for comparison, although the number of observations for them is much smaller and does not meet the above criterion. We end up with a sample of nine ULXs, including two accreting pulsars, NGC 5907 ULX-1 and NGC 7793 P13, and an intermediate-mass black hole (IMBH) candidate, ESO 243-49 HLX-1.

All of the observations of our interest were implemented in the photon-counting mode. We reduced the data using \textsc{heasoft} version 6.21.  New event files were generated using the task \textit{xrtpipeline} with standard quality cuts. The script \textit{xrtlccorr} was used to correct the effects caused by telescope vignetting and the point spread function. The source events were extracted from a circular region with a radius of 15 pixels centered at the source position, while the background contribution in the source aperture was estimated from a source-free region with a radius of 30 pixels. For each individual observation, we calculated its time-averaged net count rate, as one data point for further analysis.  We further discarded data, where the 0.3--10 keV signal-to-noise ratio is no more than 2$\sigma$.

We calculated the Lomb--Scargle periodogram \citep{Horne1986} for ULXs to search
for possible periodic or quasi-periodic signals.  As the periodogram depends on
the sampling, we only chose the part of the light curve that has a relatively high
sampling rate and discarded a small fraction of data points largely separated
in time, inclusion of which may cause an aliasing effect  but will not affect the
main conclusion.  The fraction of observations in use is 633/649 for Holmberg
(Ho) IX X-1, 572/585 for M81 X-6, 345/381 for NGC 1313 X-1, 342/378 for NGC
1313 X-2, 202/234 for NGC 4395 X-2, 371/479 for NGC 5408 X-1, 121/165 for NGC
5907 ULX-1, 82/112 for NGC 7793 P13, and 269/269 for ESO 243-49 HLX-1. The
light curves are shown in Figure~\ref{fig:lc}. The same data set was used for
another analysis. We searched periodic signals in a timescale range from 10~days to
one-half of the total time span of the selected time interval. The
periodograms for both the hard band (1--10 keV) and the soft band (0.3--1
keV), are shown in Figure~\ref{fig:pd} for nine ULXs.  The choice of 1 keV to
separate the two bands because this is roughly the boundary between the soft
and hard emission component in the X-ray spectrum for most ULXs
\citep{Sutton2013,Middleton2015}.  A 4$\sigma$ detection limit above the white
noise is shown. We did not calculate the significance due to the presence of
red noise, mainly because the relative strength of the modulation between the
soft and hard energy bands, rather than the absolute significance, matters for
the conclusion. In particular, the modulation due to precession may be
unstable and will naturally produce a less significant periodic signal with
respect to a strictly coherent signal. Plus, most of the data have been
analyzed to search for periodic signals in the literature, where the
significance has been carefully discussed.

The highest power for the hard band was found and marked in the plot; if several significant peaks or possible harmonics are shown, the one with the shortest period is chosen, because it is supported by more cycles. Then the corresponding peak in the soft band is identified for comparison.  The light curves folded at the most likely period  are shown in Figure~\ref{fig:fold} for the two bands. We emphasize that the second rule only applies to Ho IX X-1 and M81 X-6, for which no matter which peak is selected, the conclusion remains: stronger modulation in the hard band than in the soft band.

\begin{figure*}
\includegraphics[width=0.3\textwidth]{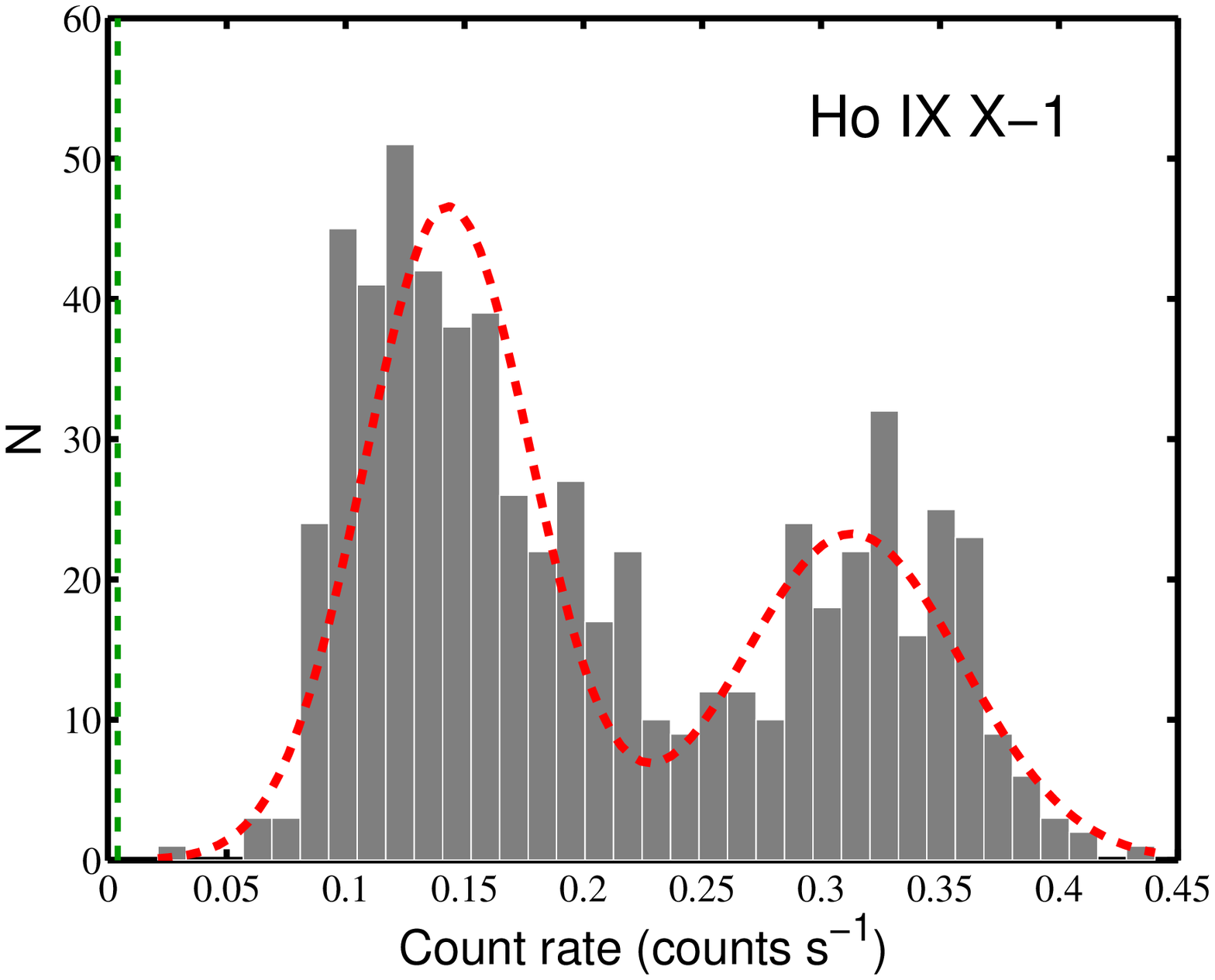}
\includegraphics[width=0.3\textwidth]{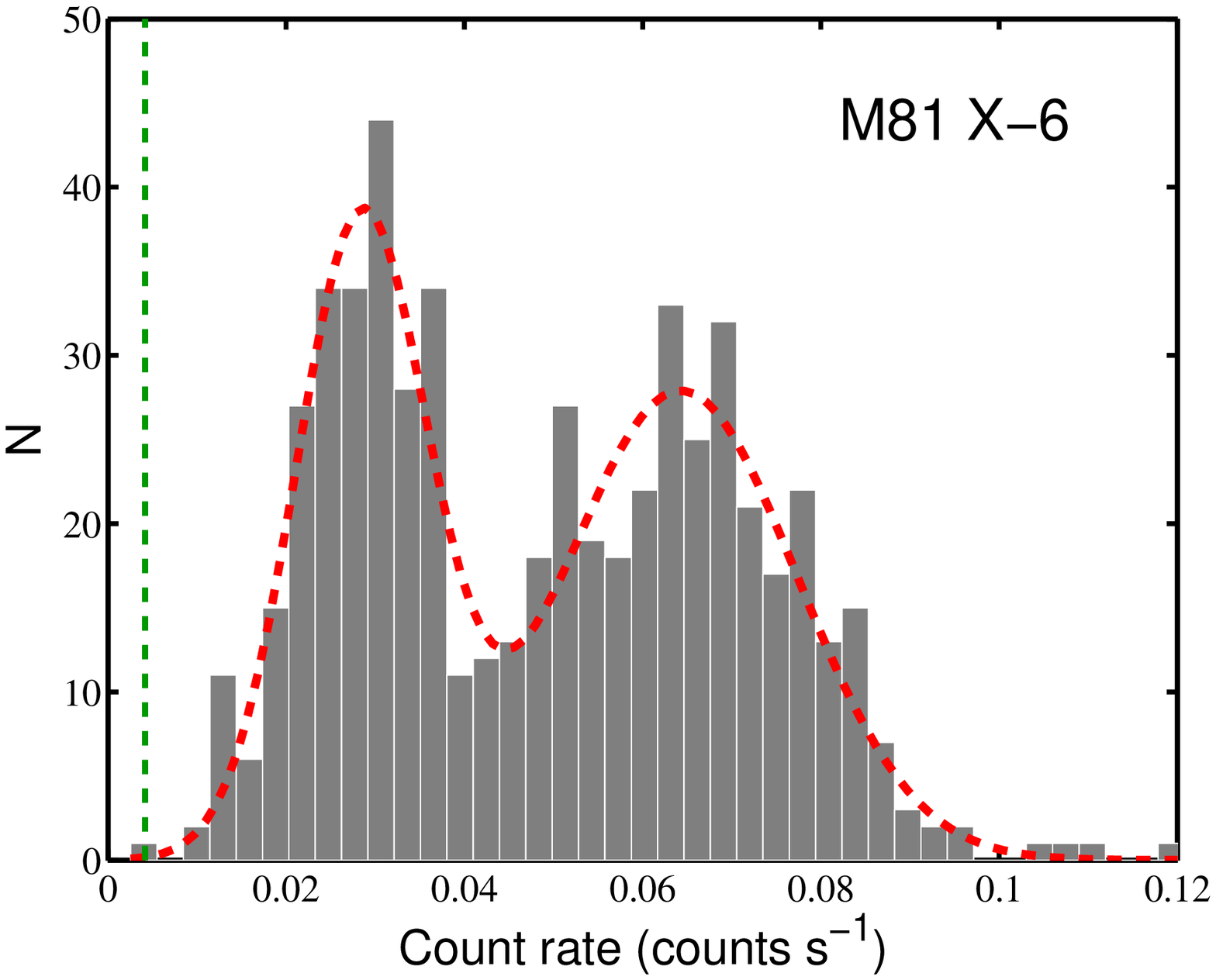}
\includegraphics[width=0.3\textwidth]{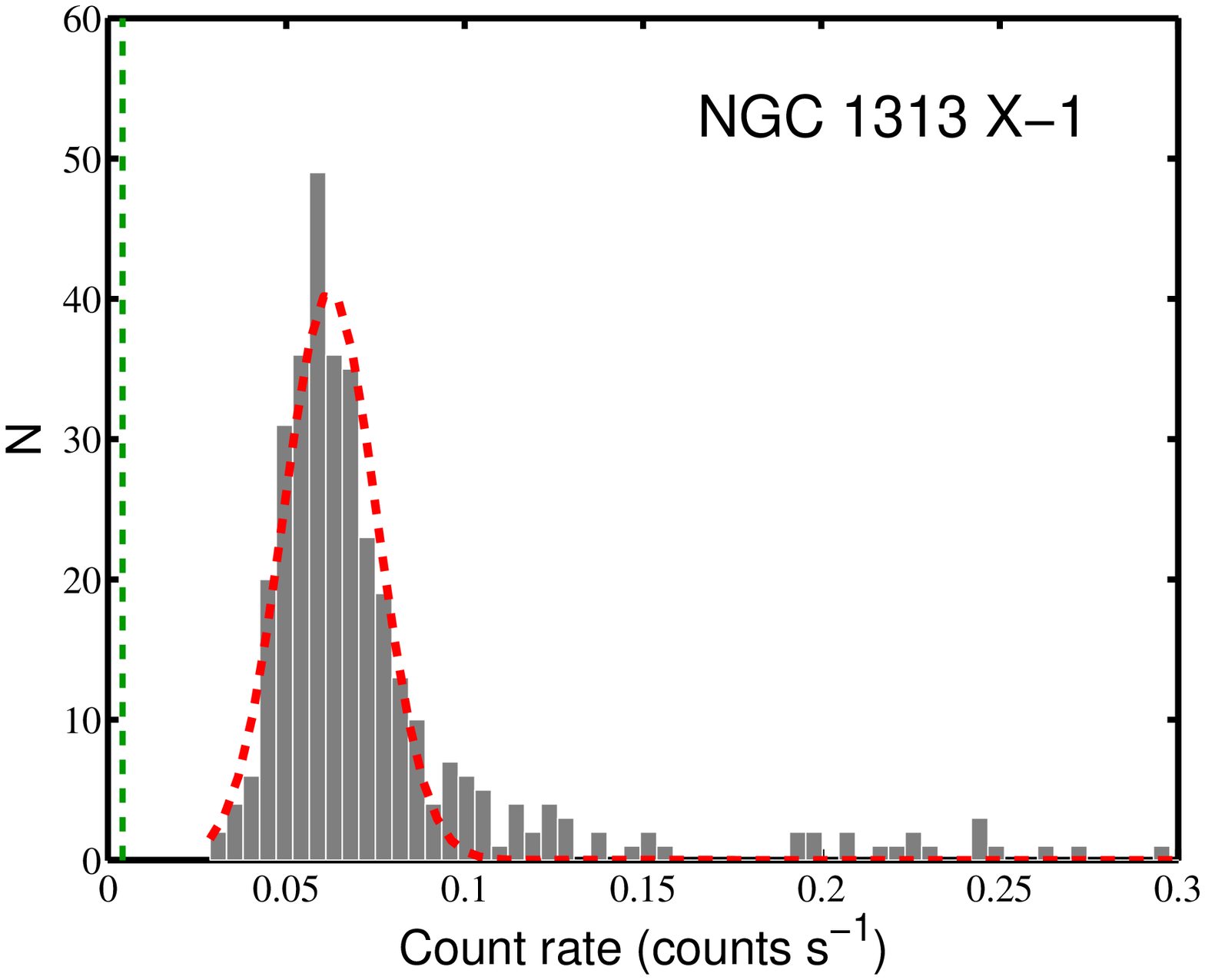}\\
\includegraphics[width=0.3\textwidth]{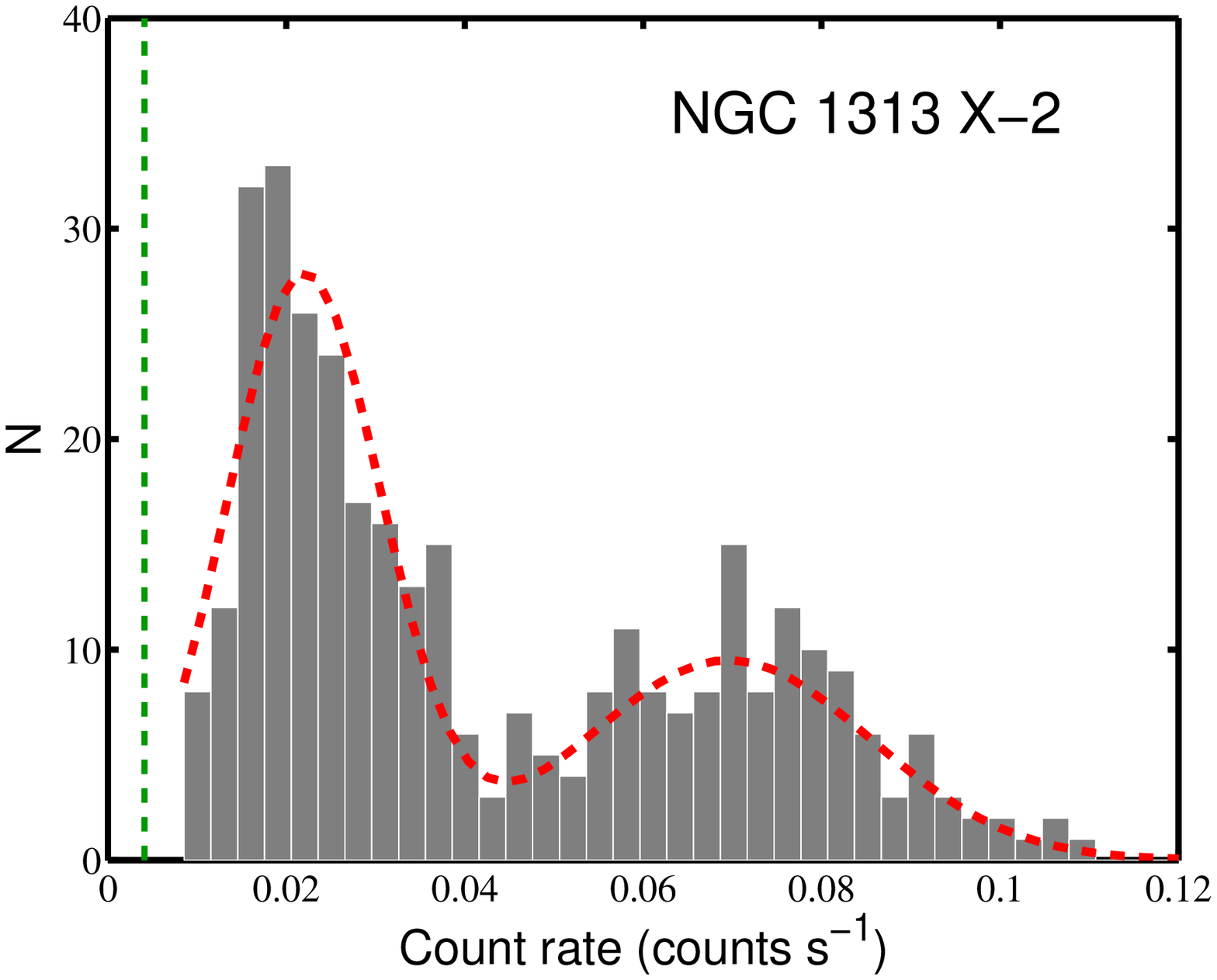}
\includegraphics[width=0.3\textwidth]{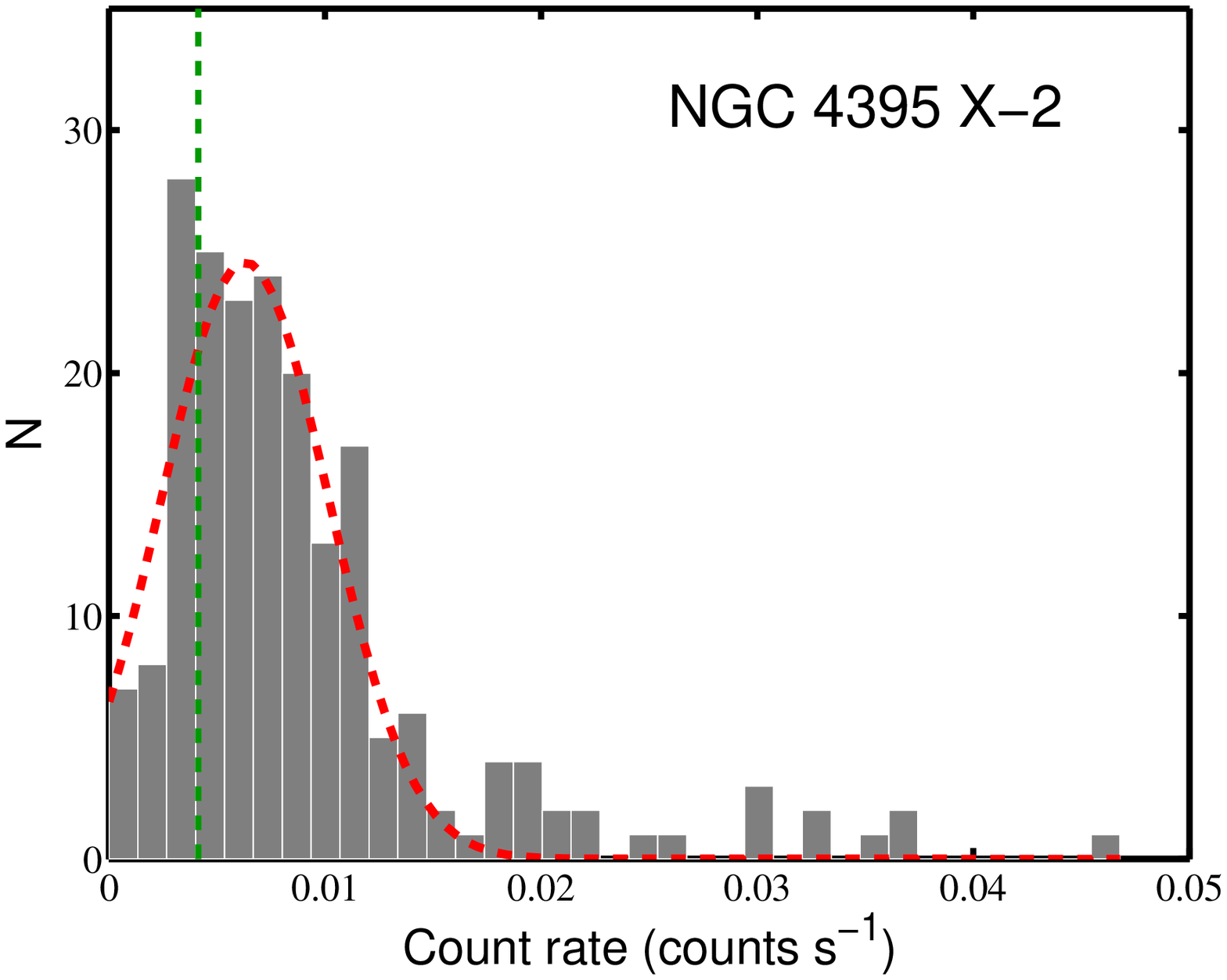}
\includegraphics[width=0.3\textwidth]{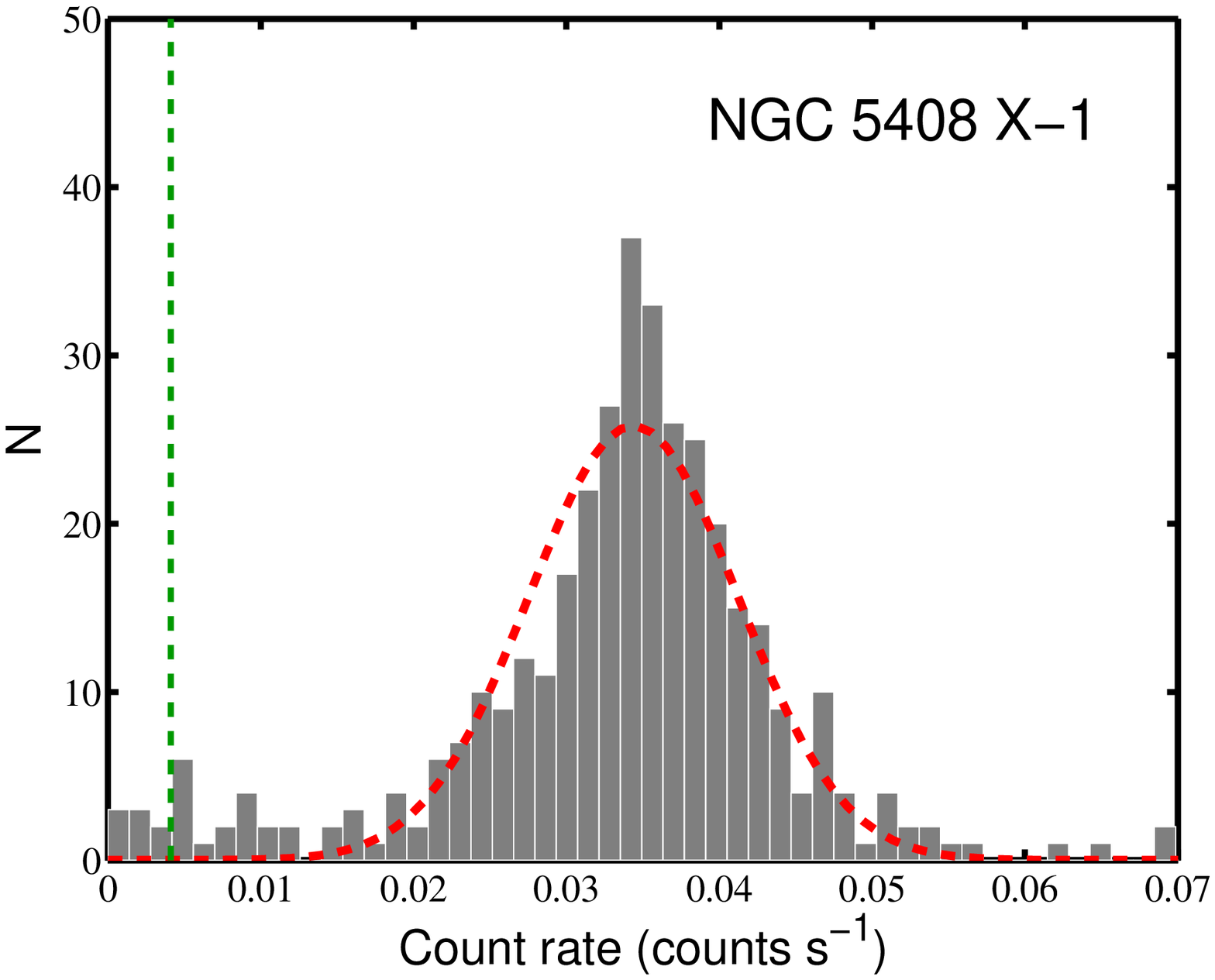}\\
\includegraphics[width=0.3\textwidth]{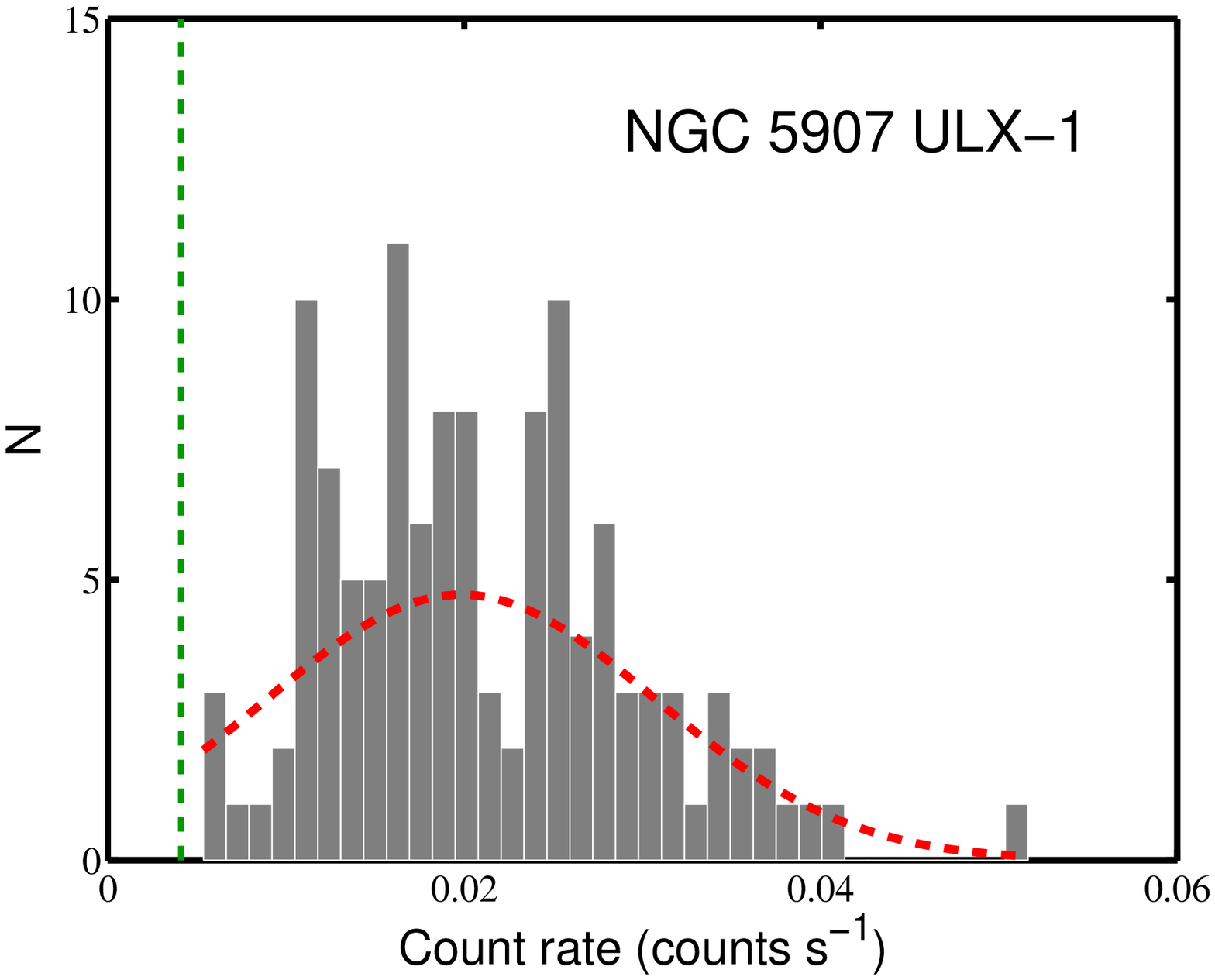}
\includegraphics[width=0.3\textwidth]{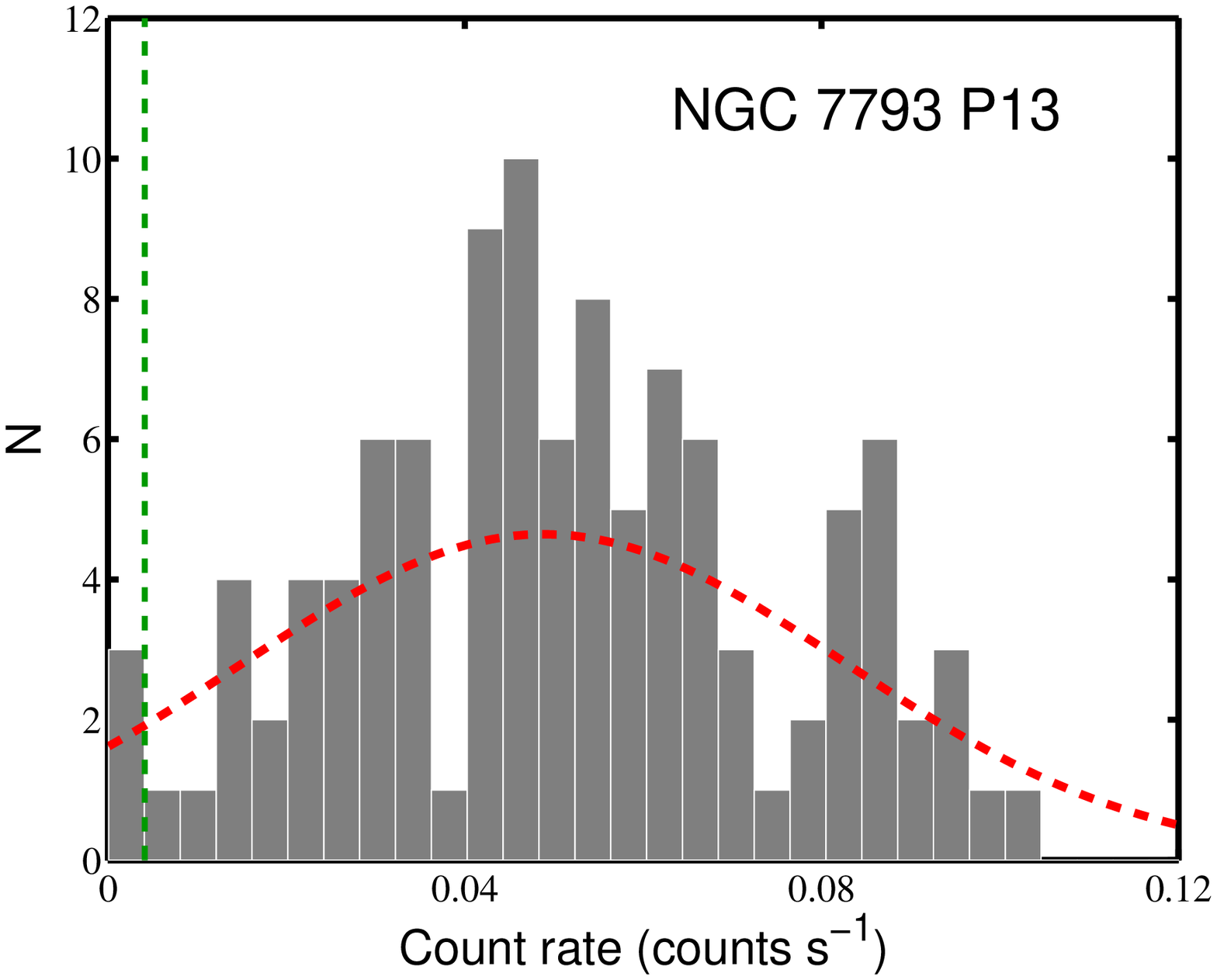}
\includegraphics[width=0.3\textwidth]{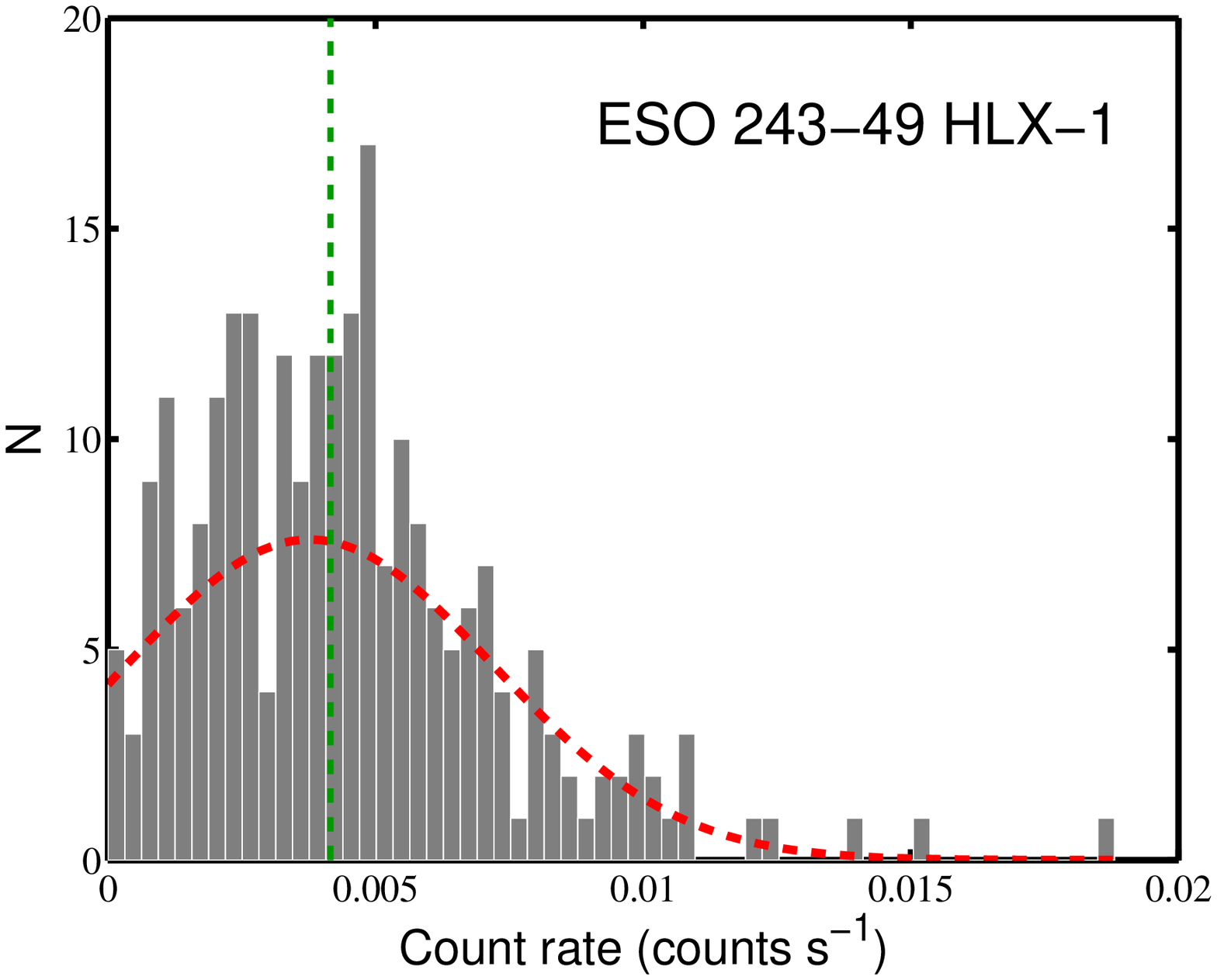}\\
\caption{Distributions of the {\it Swift} XRT 1--10 keV count rate for ULXs in the sample. Each histogram is fitted by a Gaussian function or two if needed. The green dashed line marks the 2$\sigma$ detection limit for a 1000 s observation.
\label{fig:hist_hard}}
\end{figure*}

\begin{figure*}
\includegraphics[width=0.3\textwidth]{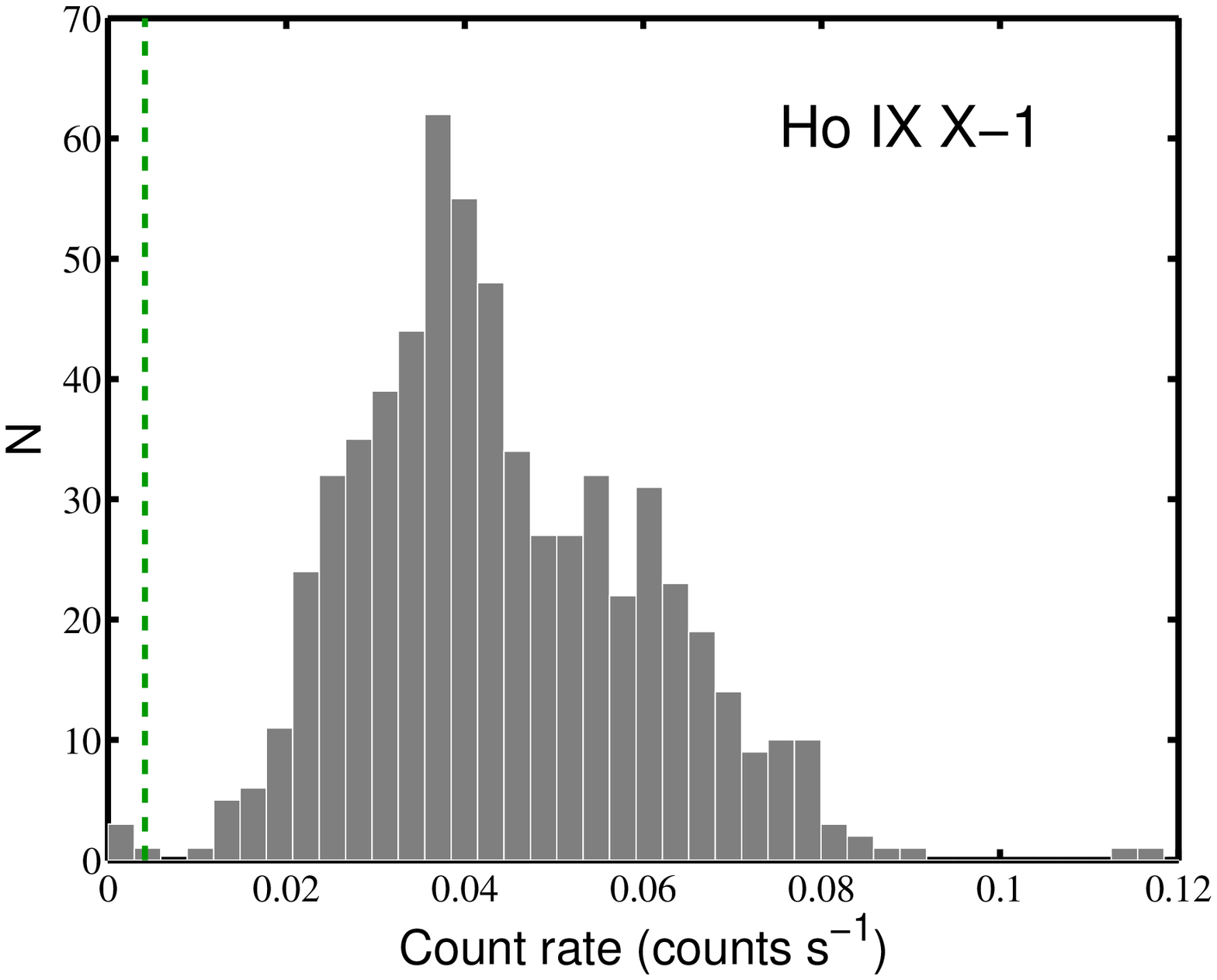}
\includegraphics[width=0.3\textwidth]{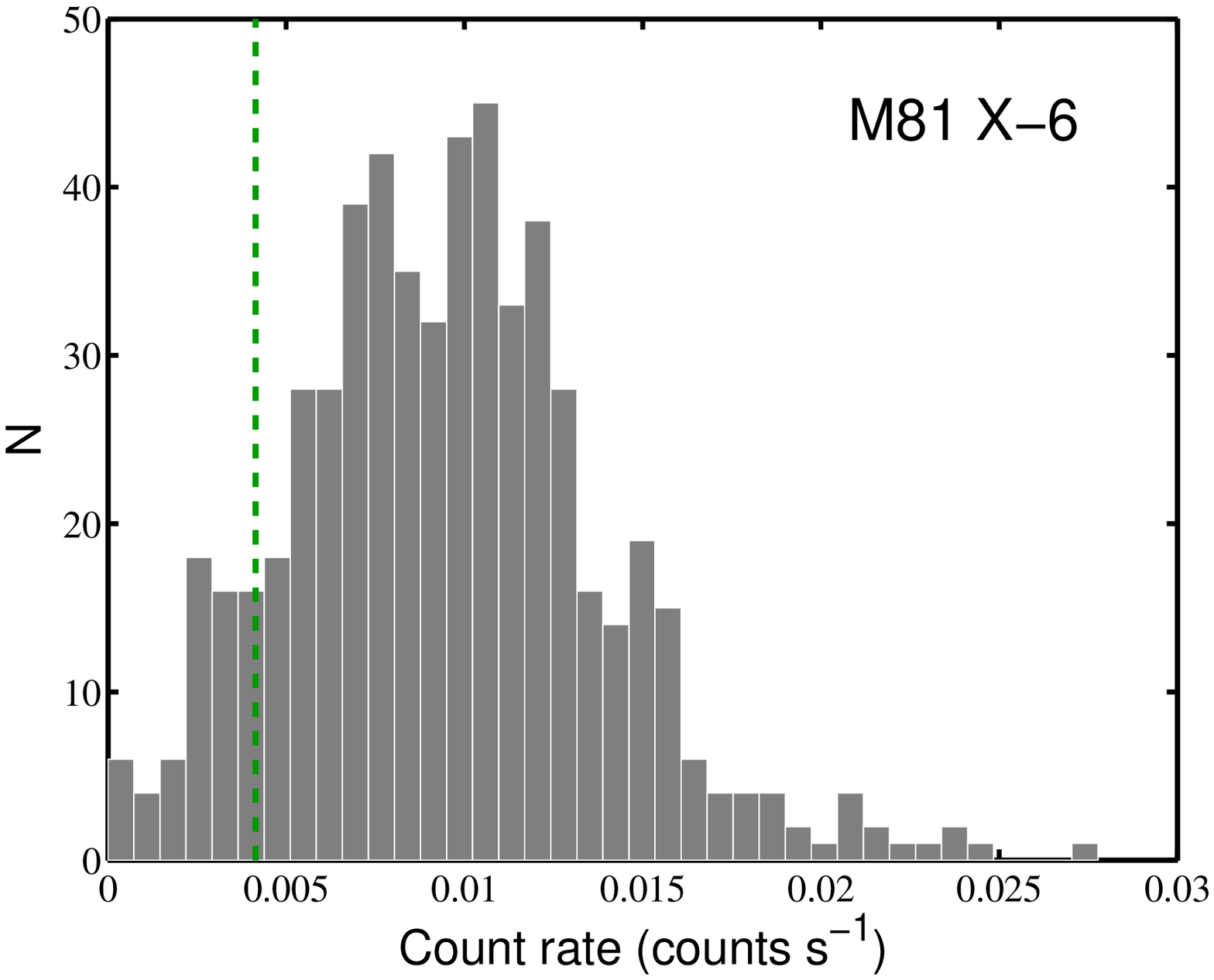}
\includegraphics[width=0.3\textwidth]{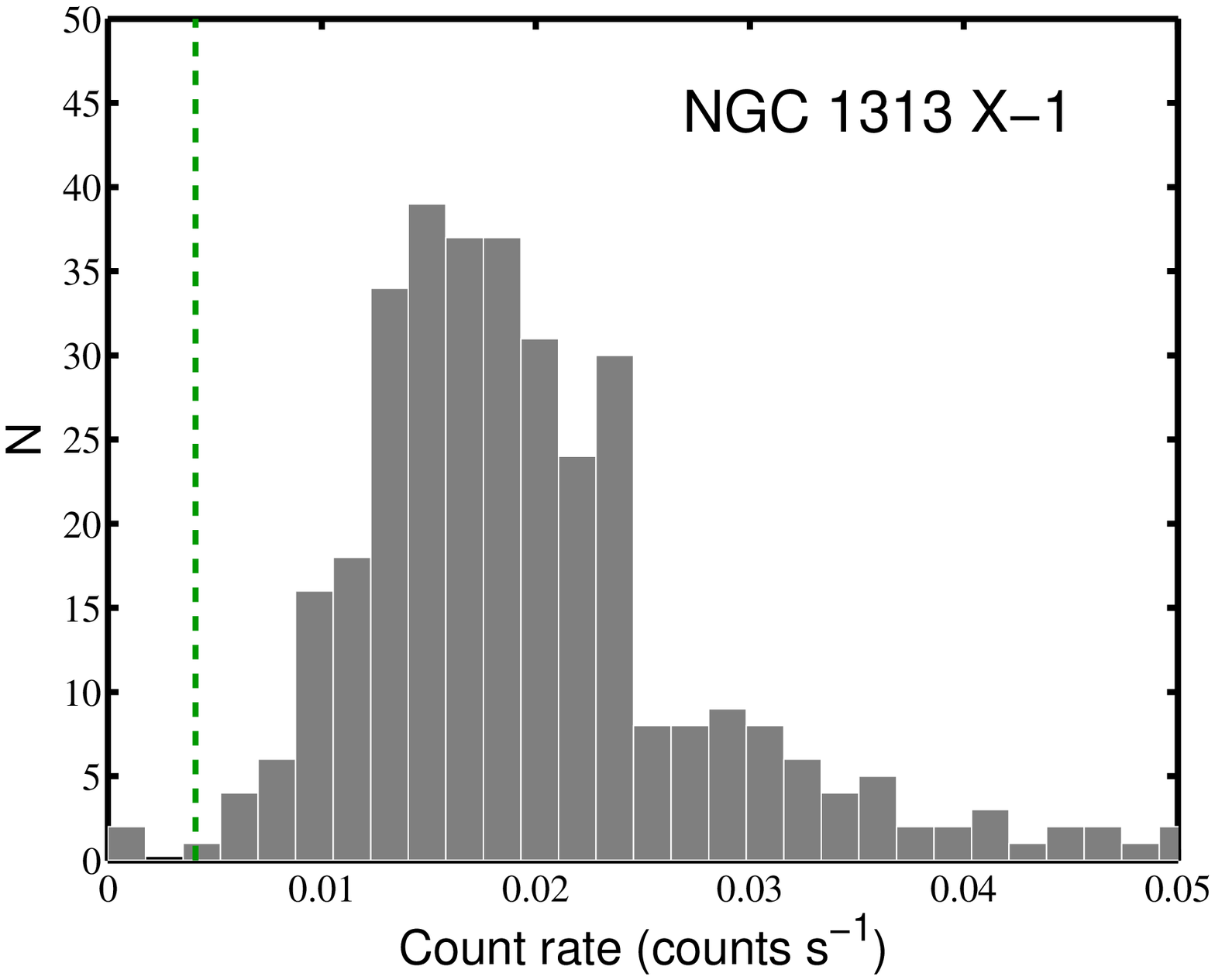}\\
\includegraphics[width=0.3\textwidth]{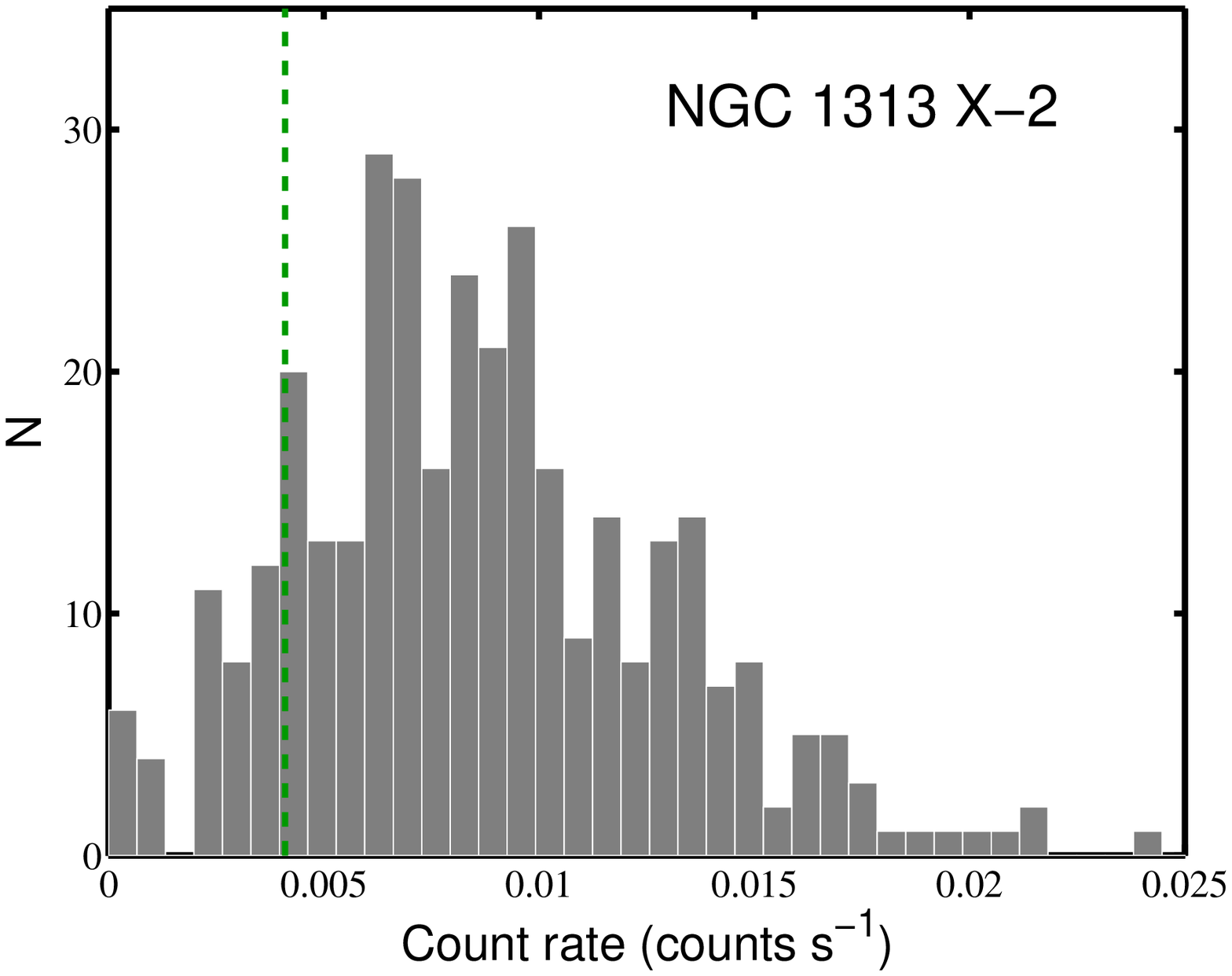}
\includegraphics[width=0.3\textwidth]{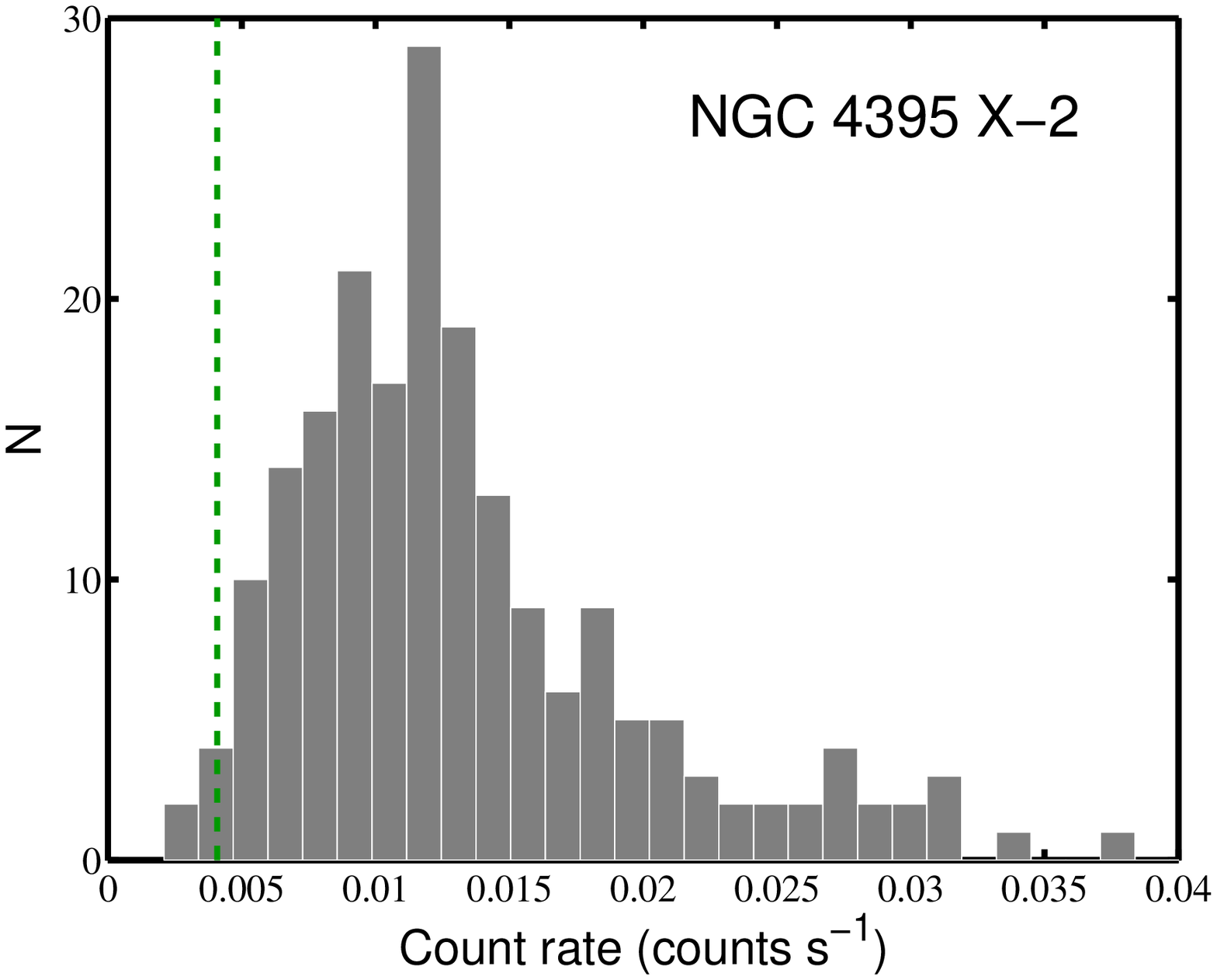}
\includegraphics[width=0.3\textwidth]{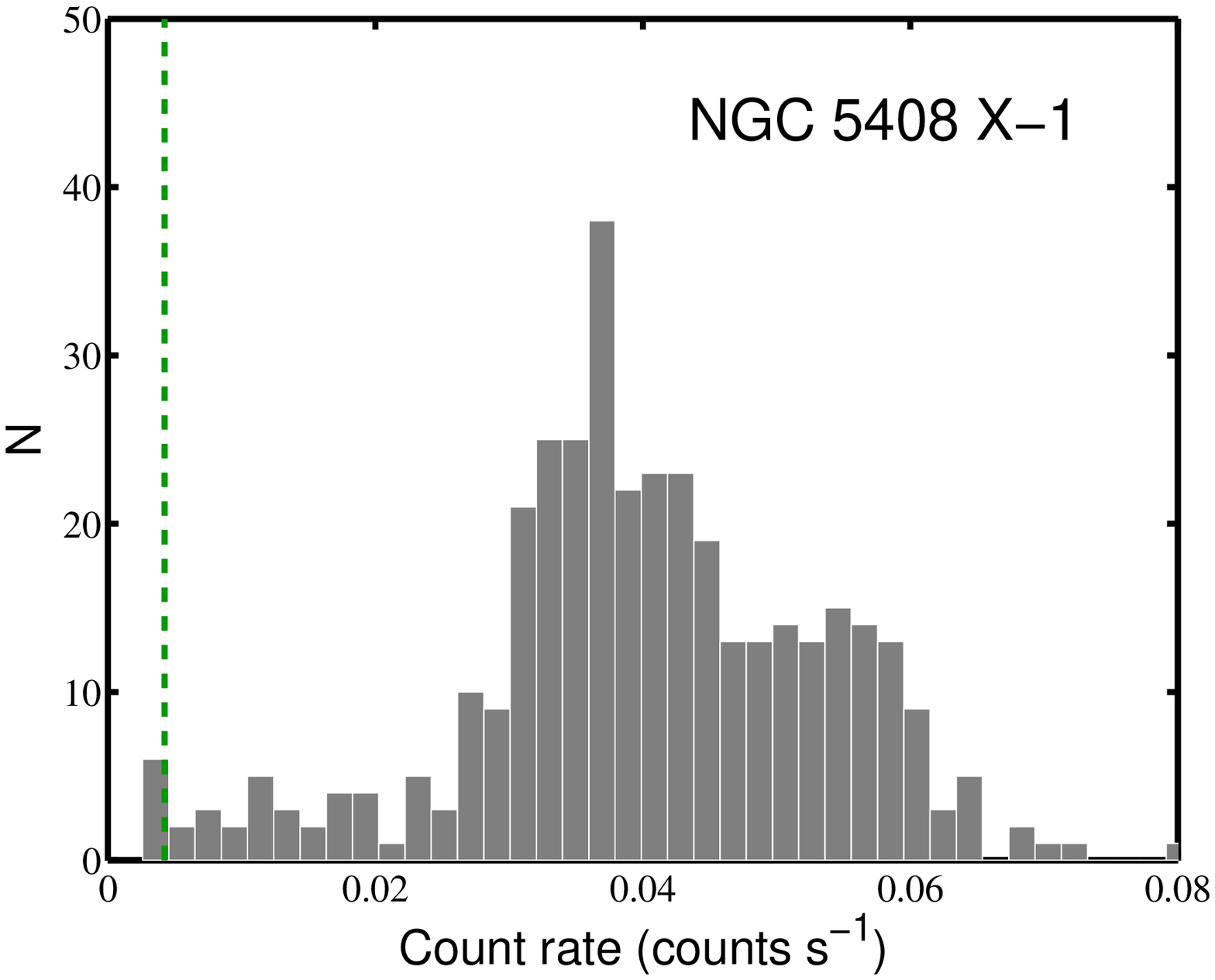}\\
\includegraphics[width=0.3\textwidth]{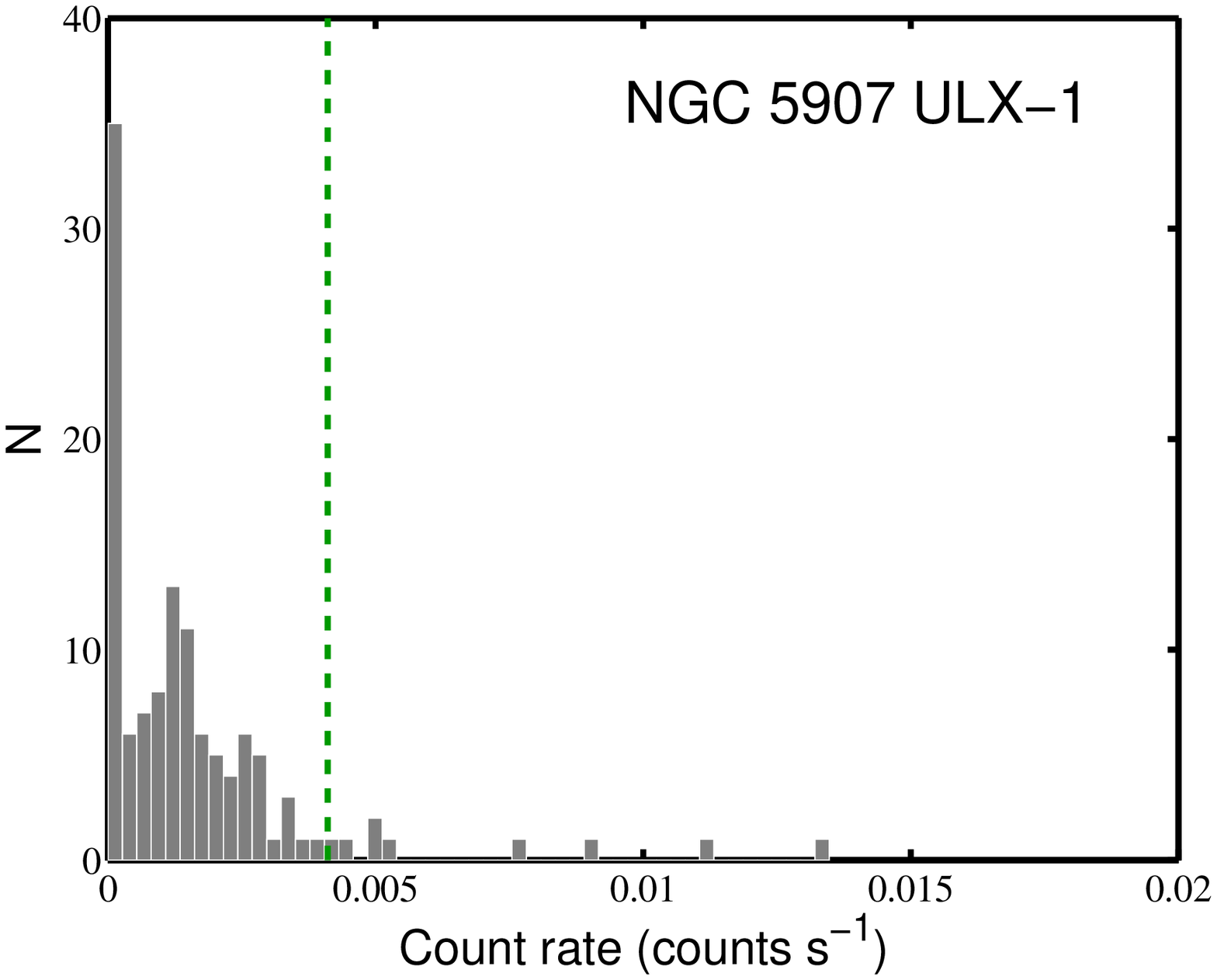}
\includegraphics[width=0.3\textwidth]{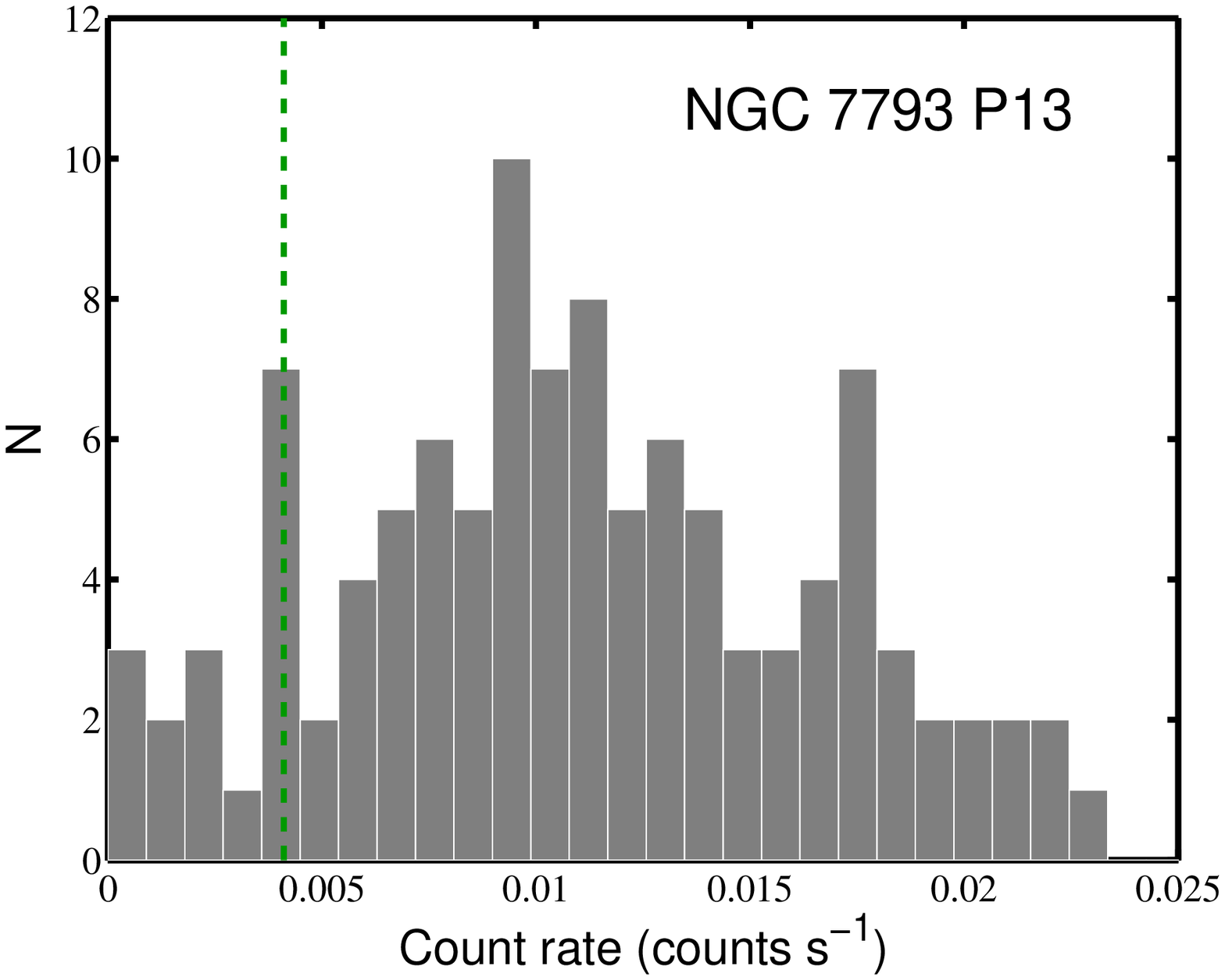}
\includegraphics[width=0.3\textwidth]{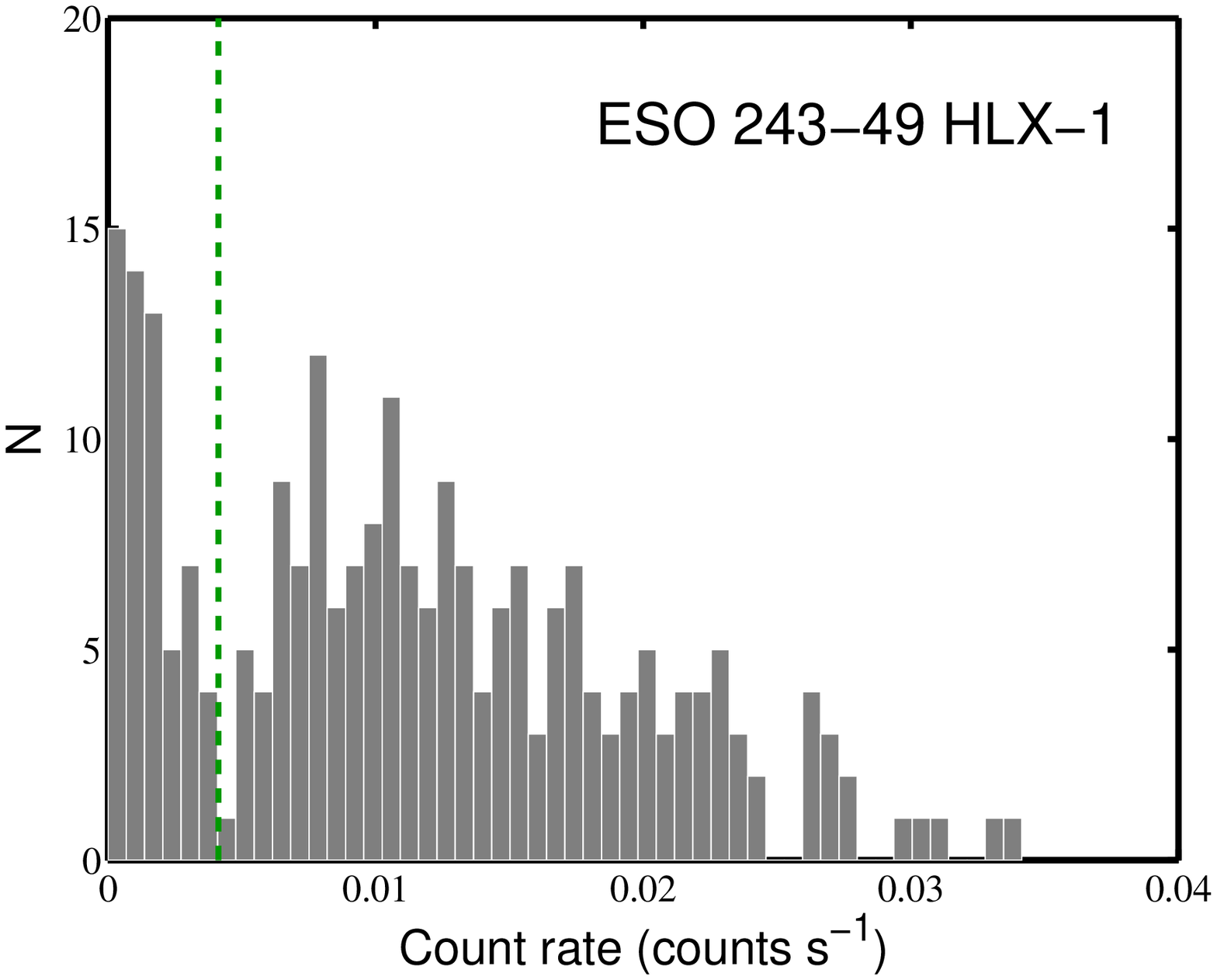}\\
\caption{Same as Figure~\ref{fig:hist_hard} but in the energy band of 0.3--1 keV.
\label{fig:hist_soft}}
\end{figure*}

The observed flux (count rate) distribution for each ULX in 1--10 keV is shown in Figure~\ref{fig:hist_hard}.  Some sources show a bimodal distribution, while others do not. We fitted the flux distribution with Gaussians and the errors for small counts were treated using the statistic in \citet[][$1+\sqrt{N+0.75}$ for $N < 5$, or $\sqrt{N}$  otherwise]{Gehrels1986}. F-test indicates that the second Gaussian component has a chance probability of $1 \times 10^{-7}$, $7 \times 10^{-12}$, $2 \times 10^{-22}$, respectively, for Ho IX X-1, M81 X-6, and NGC 1313 X-2. On the other hand, for the rest of ULXs, a single Gaussian can adequately fit the flux distribution. On each plot, the dashed green line marks the 2$\sigma$ detection limit given a typical {\it Swift} exposure of 1~ks. This suggests that the lower-flux peak in the bimodal distribution is not an observational effect. The flux distributions in the 0.3--1 keV energy band is shown in Figure~\ref{fig:hist_soft}. None of the ULXs shows a bimodal distribution in the soft band.

In Figure~\ref{fig:hid}, we plotted the ratio of count rates between the hard and soft bands (1--10 keV/0.3--1 keV) versus the observed XRT count rate in 0.3--10 keV, \textit{a.k.a.} the hardness intensity diagram (HID).  As one can see, the bimodal distribution is even more obvious on the two-dimensional plane; the same three ULXs show two islands of points on the HID, corresponding to a low/soft and a high/hard regime, respectively. NGC 1313 X-1 also shows a possible bimodal distribution, but not significant enough due to a small number of points in the high/hard regime. We emphasize that both the intensity and the hardness are direct observables with {\it Swift} XRT. If absorption corrected quantities are used, the two peaks will be more distinct.

We summarize the above results in Table~\ref{tab:summary}, including whether or not they show a bimodal distribution on the HID, the luminosity ratio of the two peaks for the bimodal feature, the significance of the power in the soft and hard bands (\textit{moderate} means close to or slightly above 4$\sigma$, otherwise being \textit{strong} or \textit{weak}), the peak period if the hard X-ray periodicity is strong or moderate, and the power ratio at the peak between the hard and soft bands.

\begin{deluxetable}{lclll}
\tablecolumns{5}
\tablewidth{0pc}
\tablecaption{Summary of the properties of ULXs in the sample.
\label{tab:summary}}
\tablehead{
\colhead{Source} & \colhead{Bimodal on HID} &  \multicolumn{3}{c}{Periodicity}\\
\noalign{\smallskip}\cline{3-5}\noalign{\smallskip}
\colhead{} & \colhead{(Flux Ratio$^\dagger$)} & \colhead{1--10 keV} & \colhead{0.3--1 keV} & \colhead{Power Ratio}
}
\startdata
Ho IX X-1 & yes (2.2) & strong (266 days) & moderate & $52.8 / 26.9 = 1.9$ \\
M81 X-6 & yes (2.3) & strong (115 days) & weak & $31.7 / 6.2 = 5.1$ \\
NGC 1313 X-1 & likely & moderate (212 days) & weak & $17.2 / 13.7 = 1.3$ \\
NGC 1313 X-2 & yes (3.2) & strong (158 days) & weak & $30.5 / 10.4 = 2.9$ \\
NGC 4395 X-2 & no & weak & weak & $12.0 / 13.2 = 0.9$ \\
NGC 5408 X-1 & no & weak & weak & $10.9 / 5.0 = 2.2$ \\
NGC 5907 ULX-1 & no & strong (79 days) & weak & $29.2 / 3.4 = 8.7$ \\
ESO 243-49 HLX-1 & no & moderate (184 days) & strong & $16.3 / 28.7 = 0.6$ \\
\enddata
\tablenotetext{\dagger}{The flux ratio is quoted as the ratio of the count rate for the two peaks in 1-10 keV (Figure~\ref{fig:hist_hard}, and the typical statistical error is 0.1.}
\end{deluxetable}

\section{Discussion and conclusion}
\label{sec:dis}

\begin{figure*}
\includegraphics[width=0.49\textwidth]{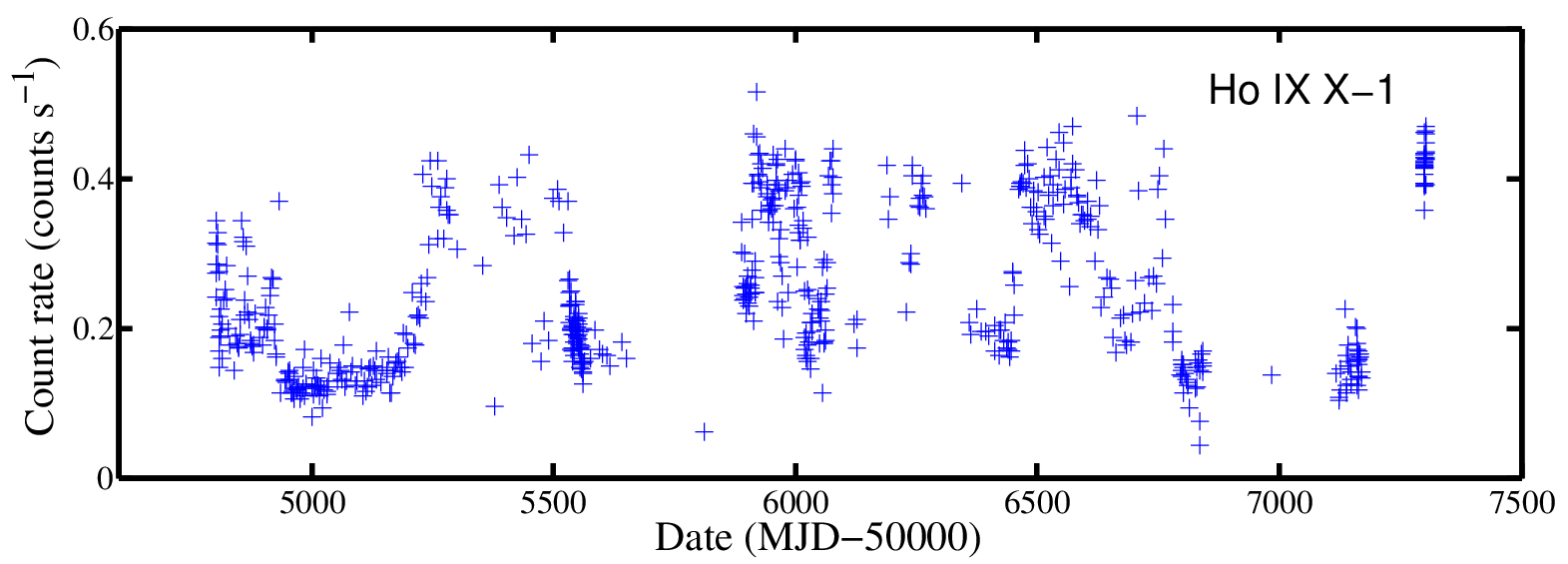}
\includegraphics[width=0.49\textwidth]{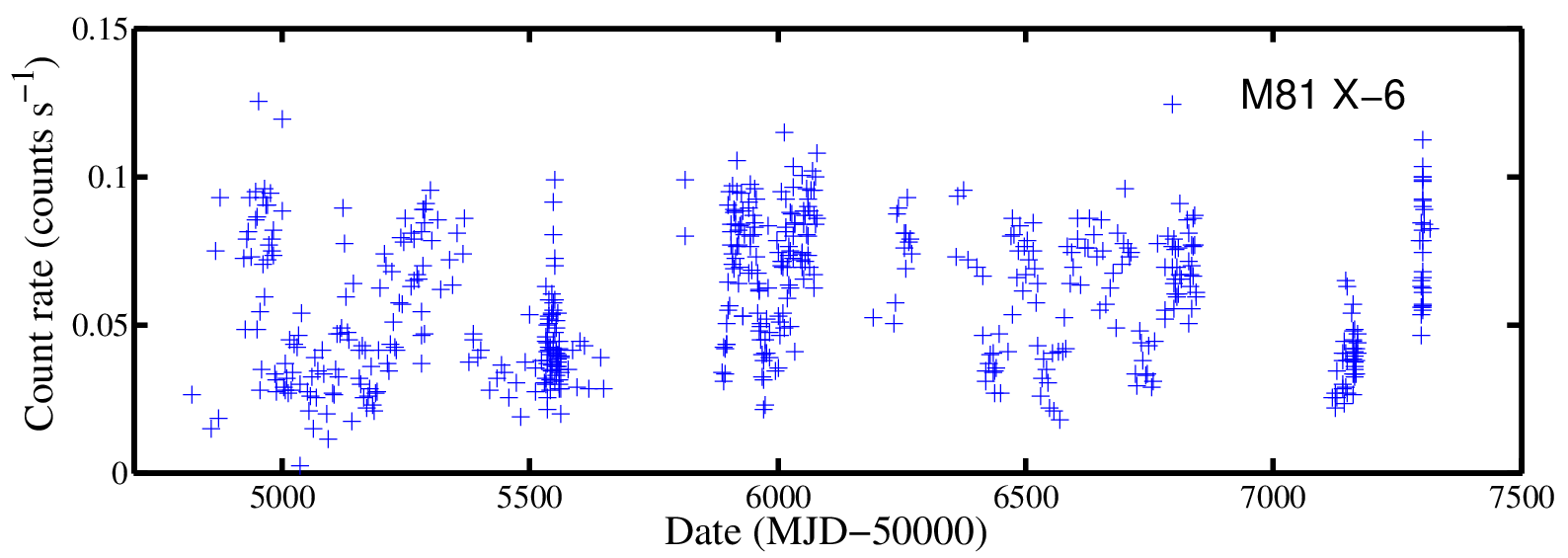}\\
\includegraphics[width=0.49\textwidth]{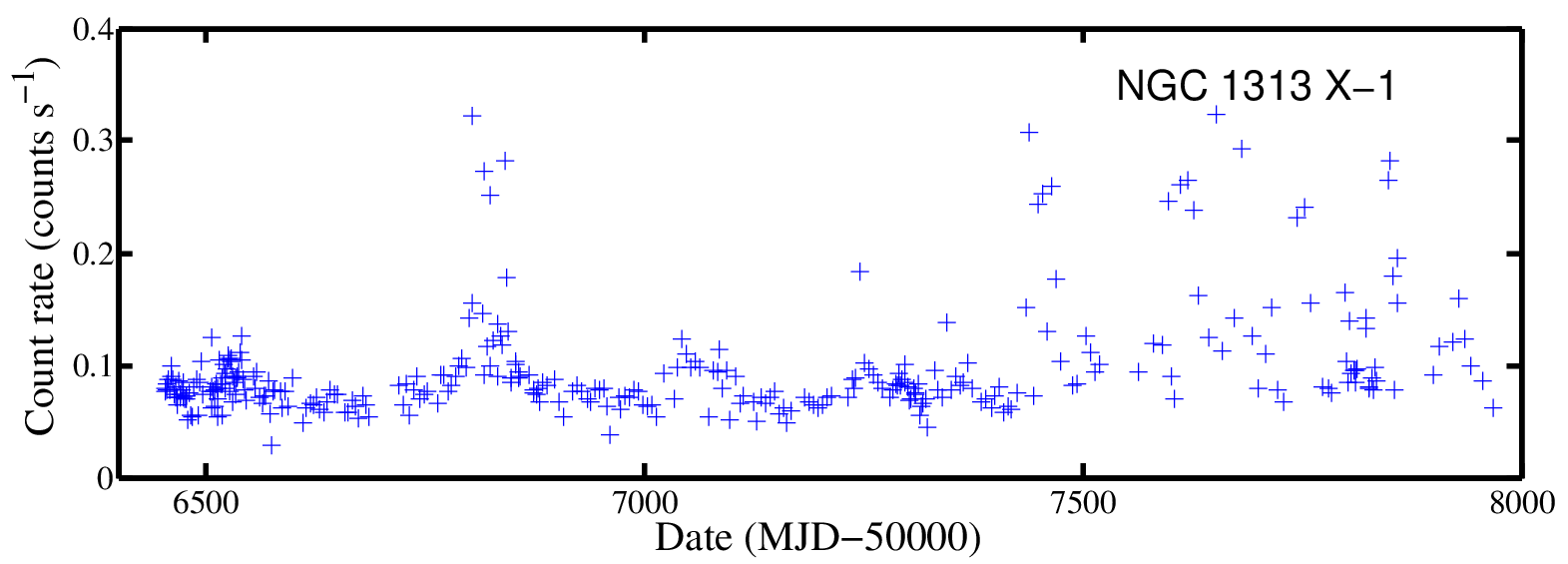}
\includegraphics[width=0.49\textwidth]{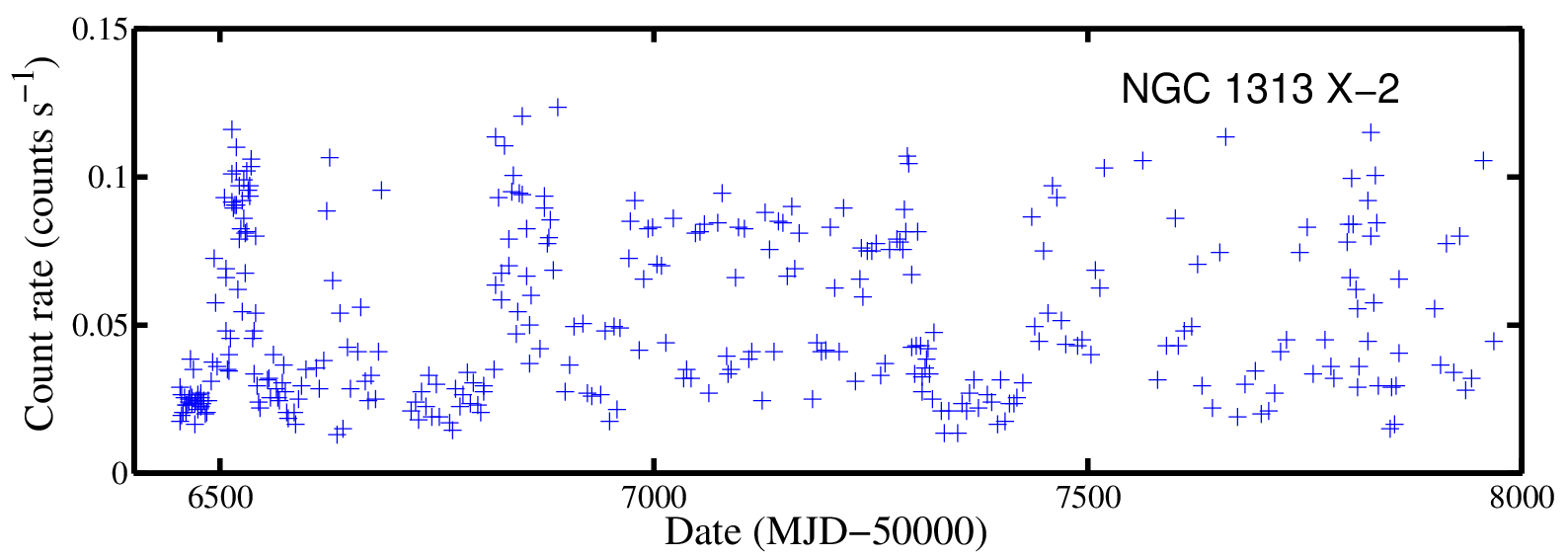}\\
\includegraphics[width=0.49\textwidth]{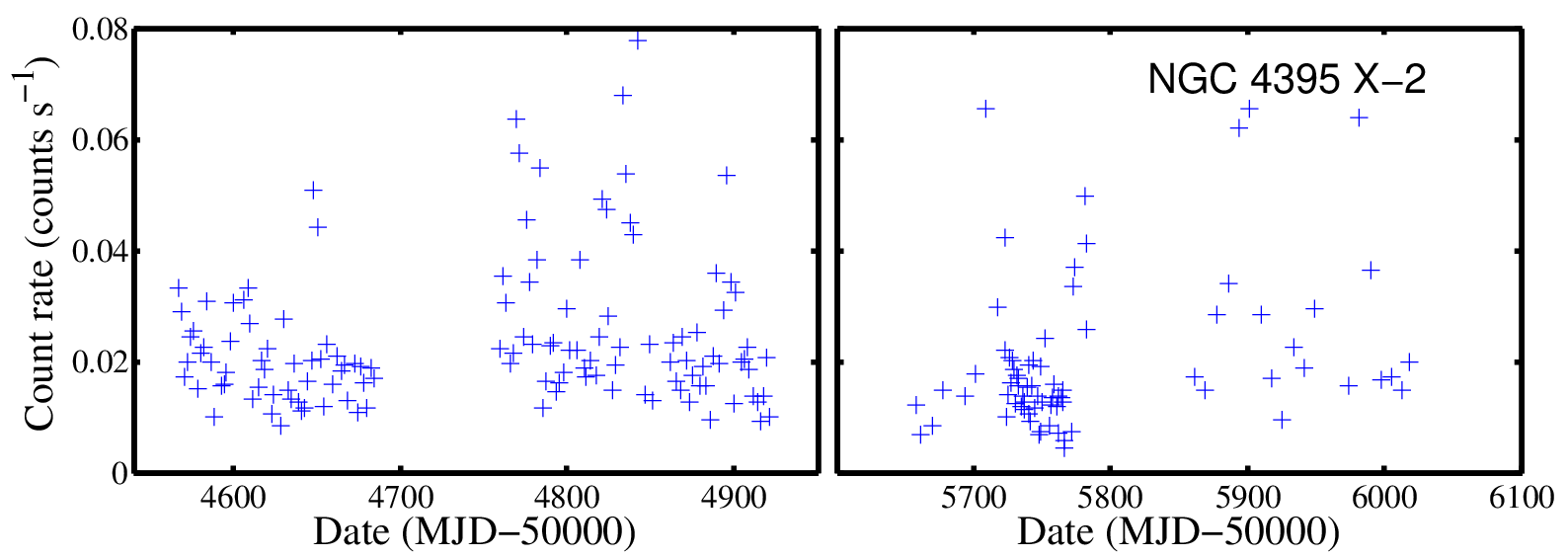}
\includegraphics[width=0.49\textwidth]{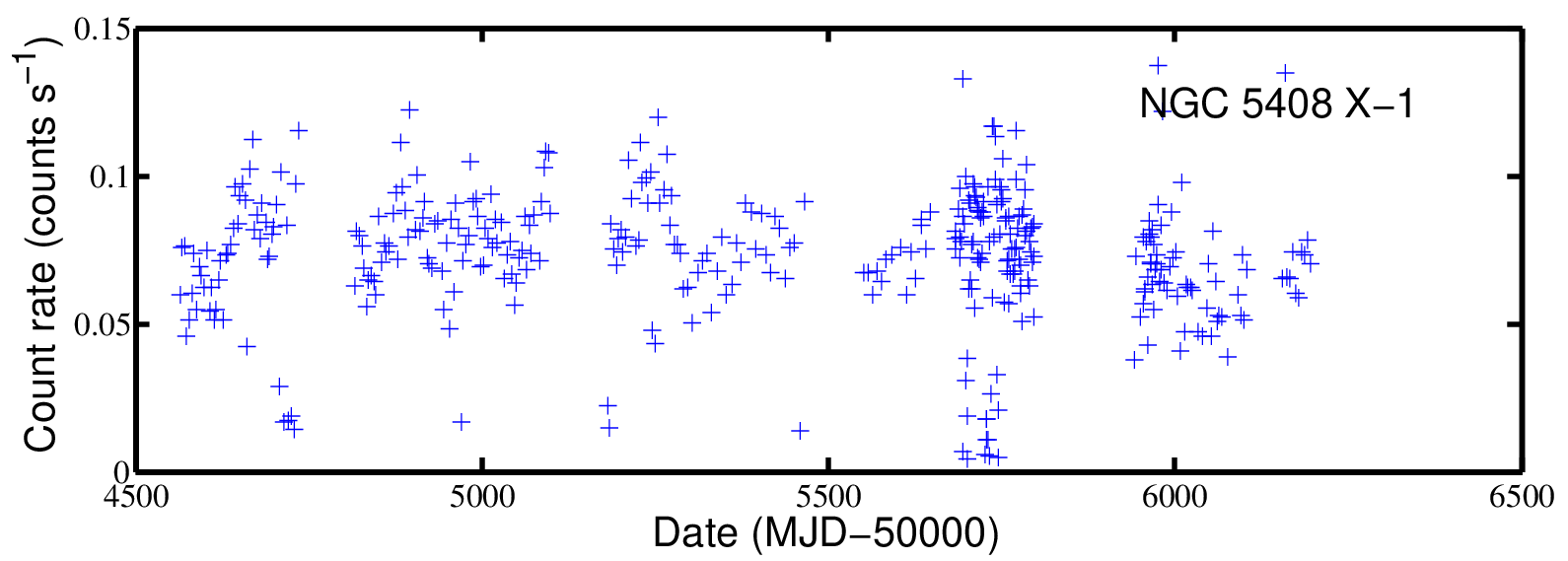}\\
\includegraphics[width=0.49\textwidth]{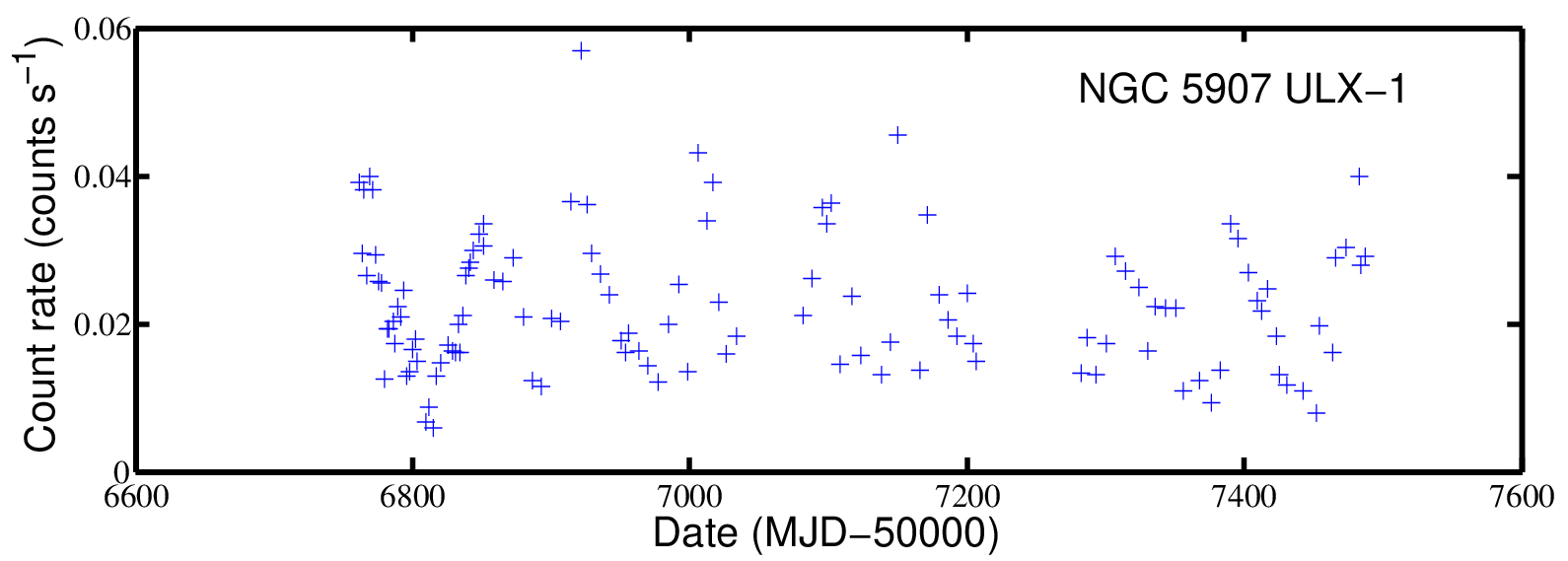}
\includegraphics[width=0.49\textwidth]{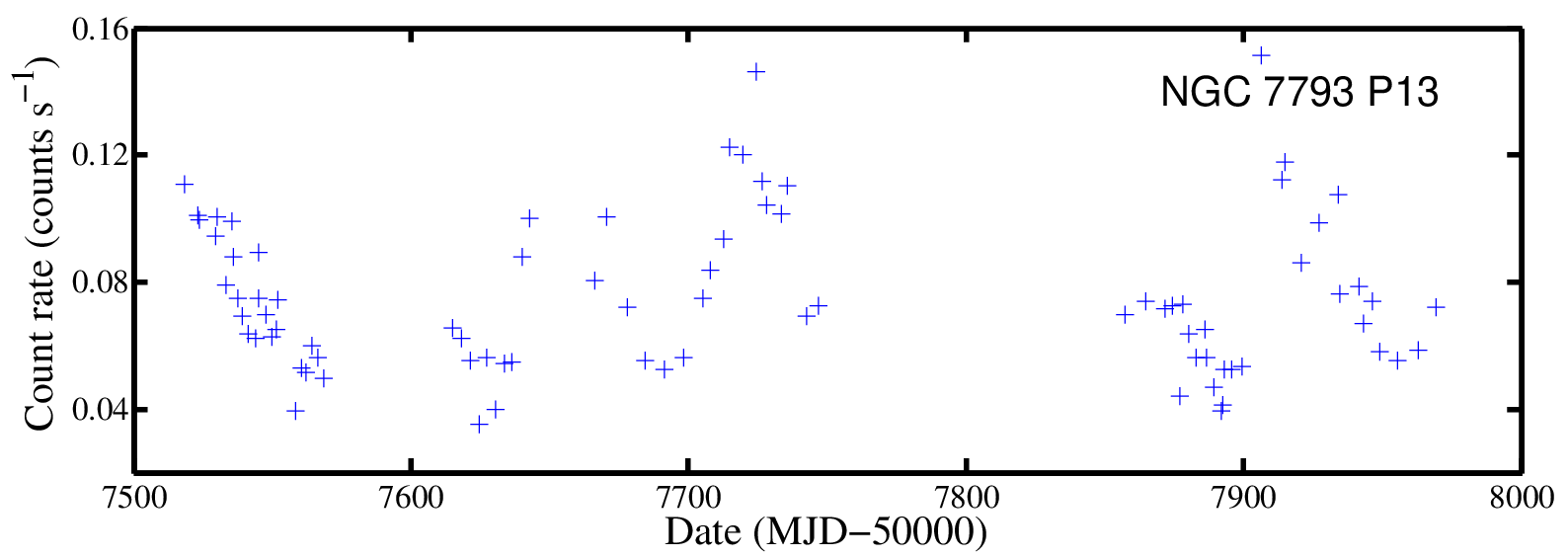}\\
\includegraphics[width=0.49\textwidth]{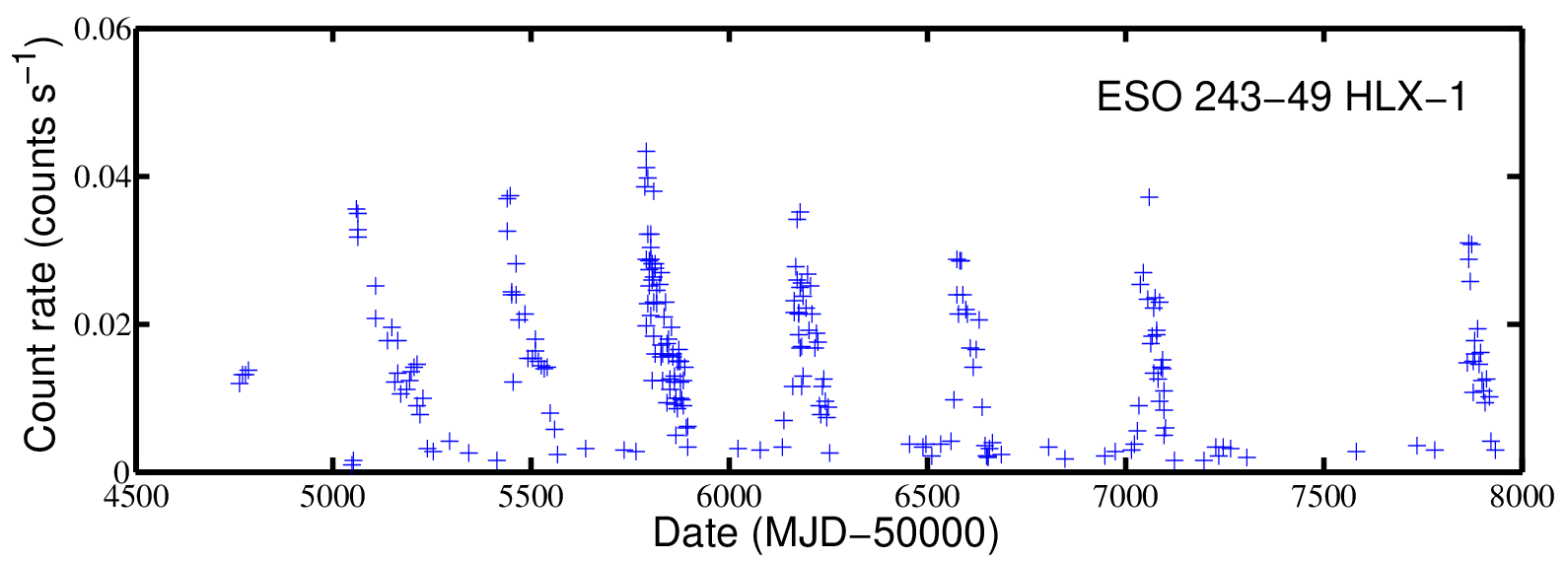}
\caption{{\it Swift}/XRT 0.3--10 keV light curves (observed count rates) for ULXs in our sample.  Due to uneven sampling, we only adopt the duration when the sampling is the densest for each source. A small fraction of data that are largely separated in time are discarded.
\label{fig:lc}}
\end{figure*}

\begin{figure*}
\includegraphics[width=0.3\textwidth]{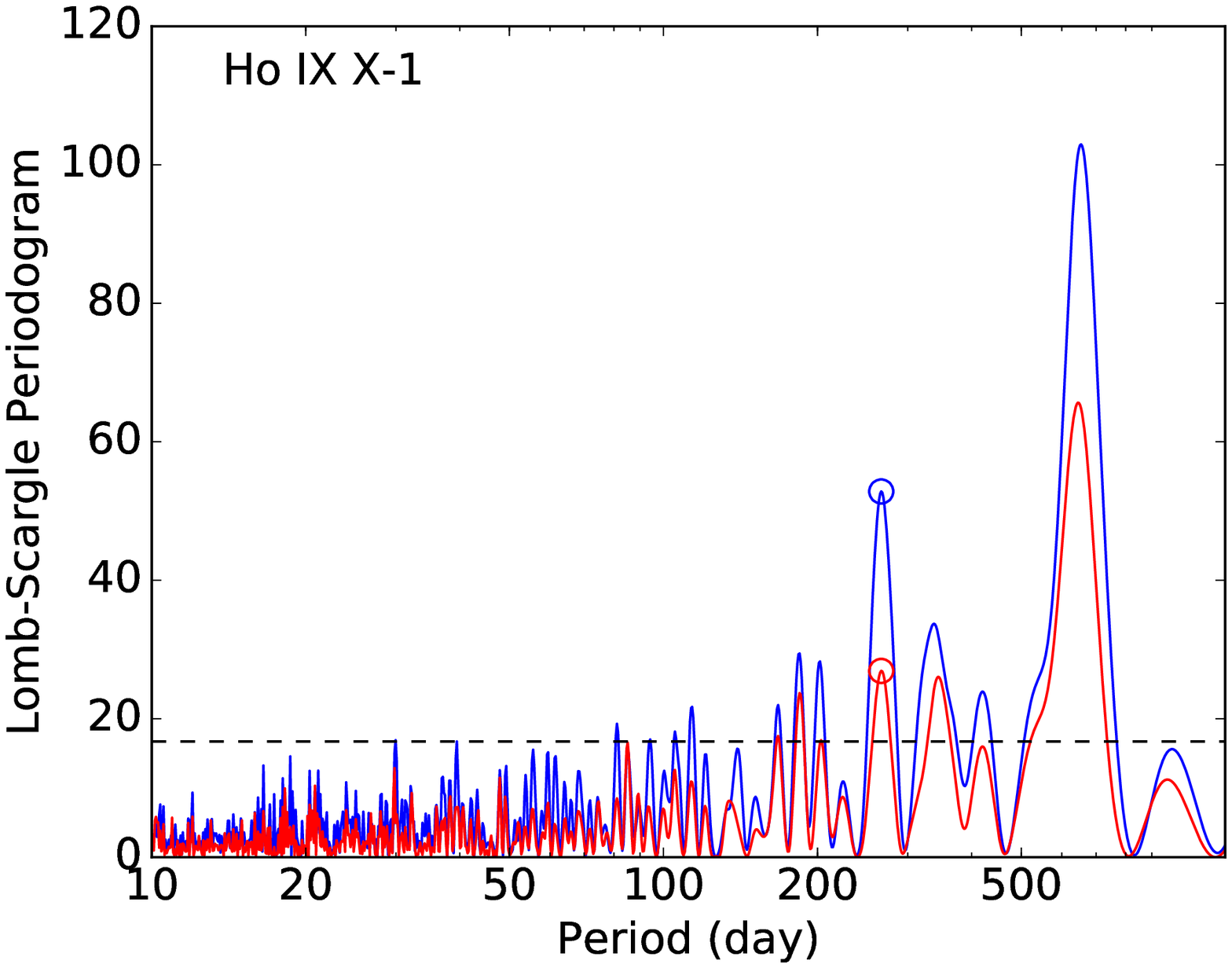}
\includegraphics[width=0.3\textwidth]{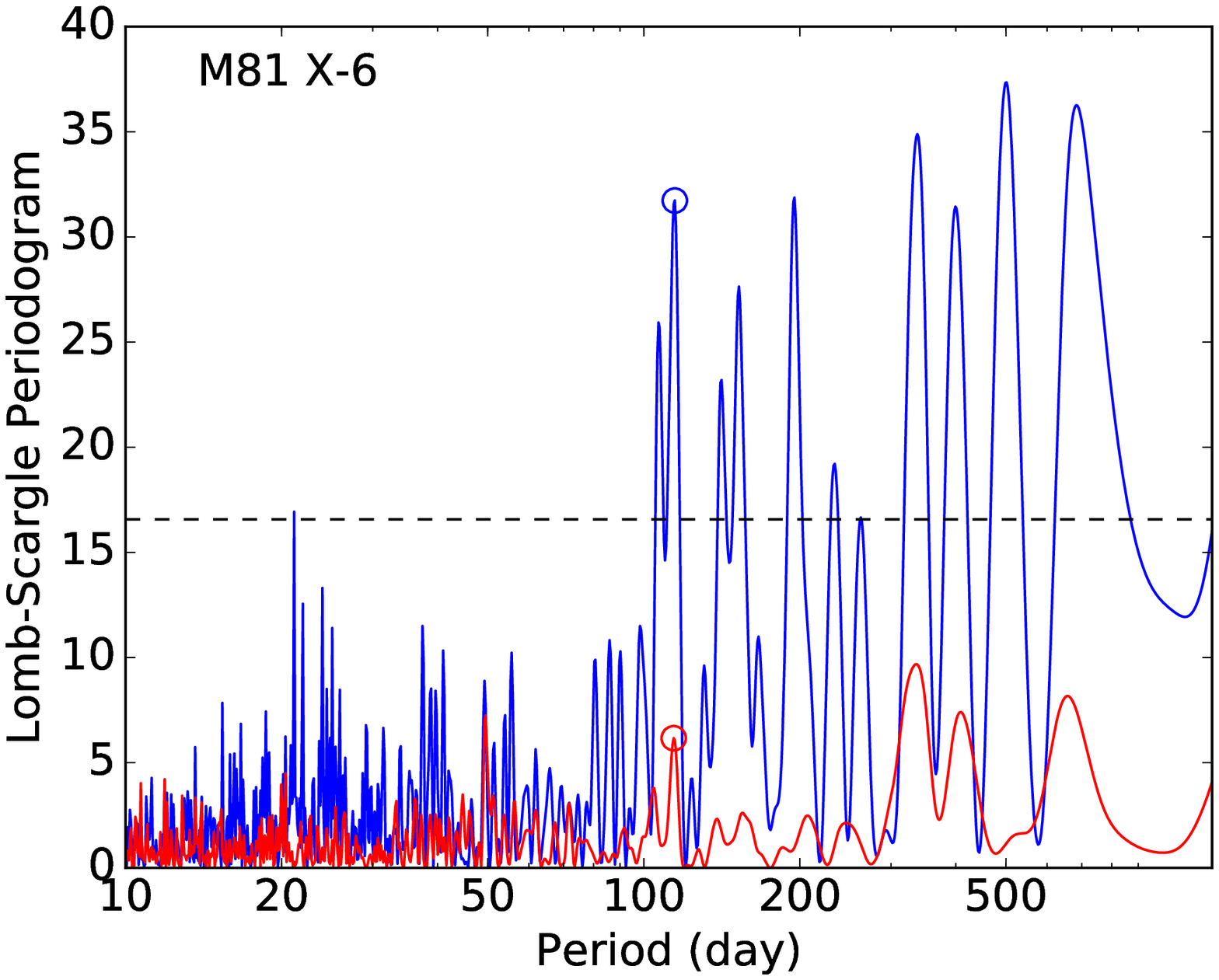}
\includegraphics[width=0.3\textwidth]{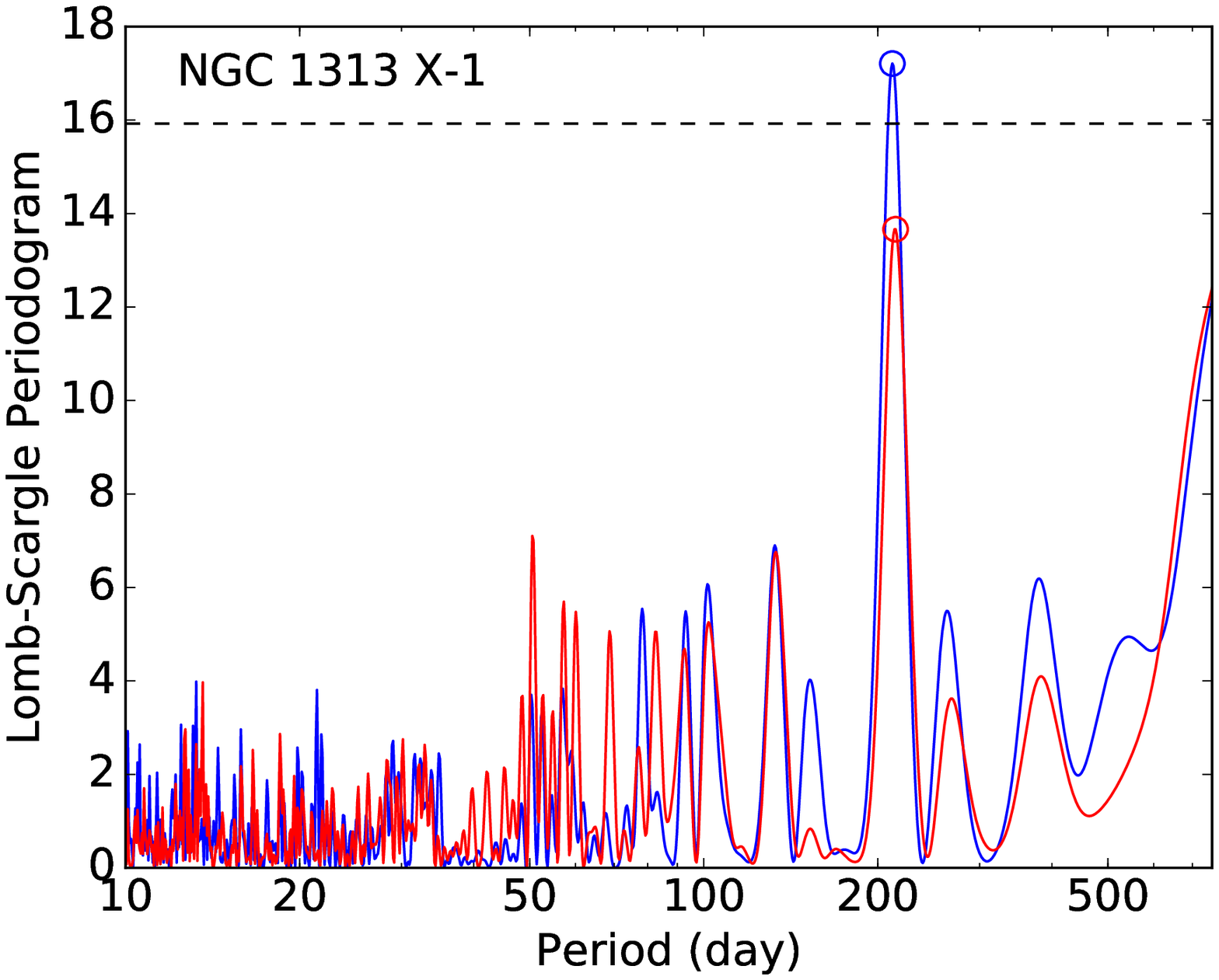}\\
\includegraphics[width=0.3\textwidth]{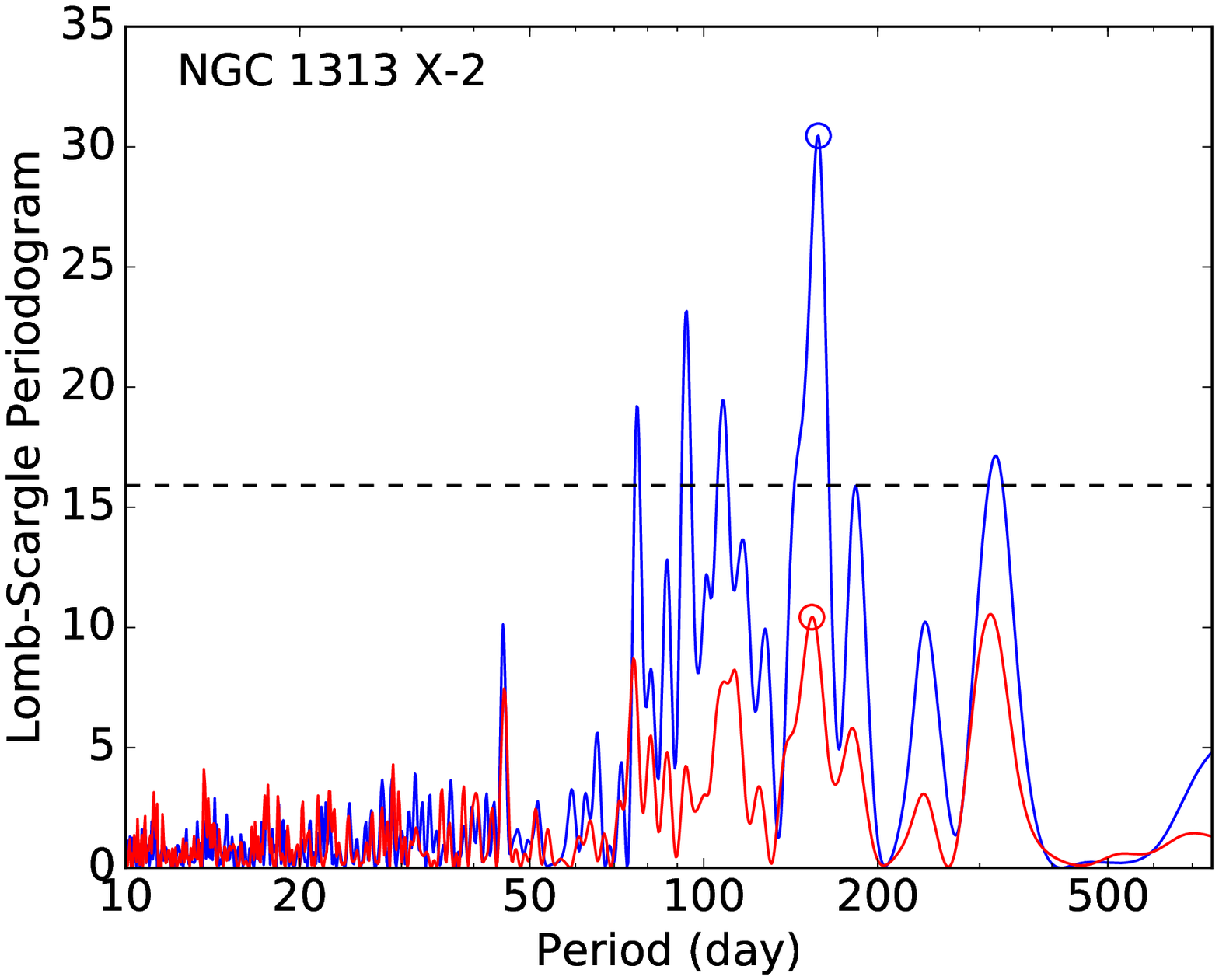}
\includegraphics[width=0.3\textwidth]{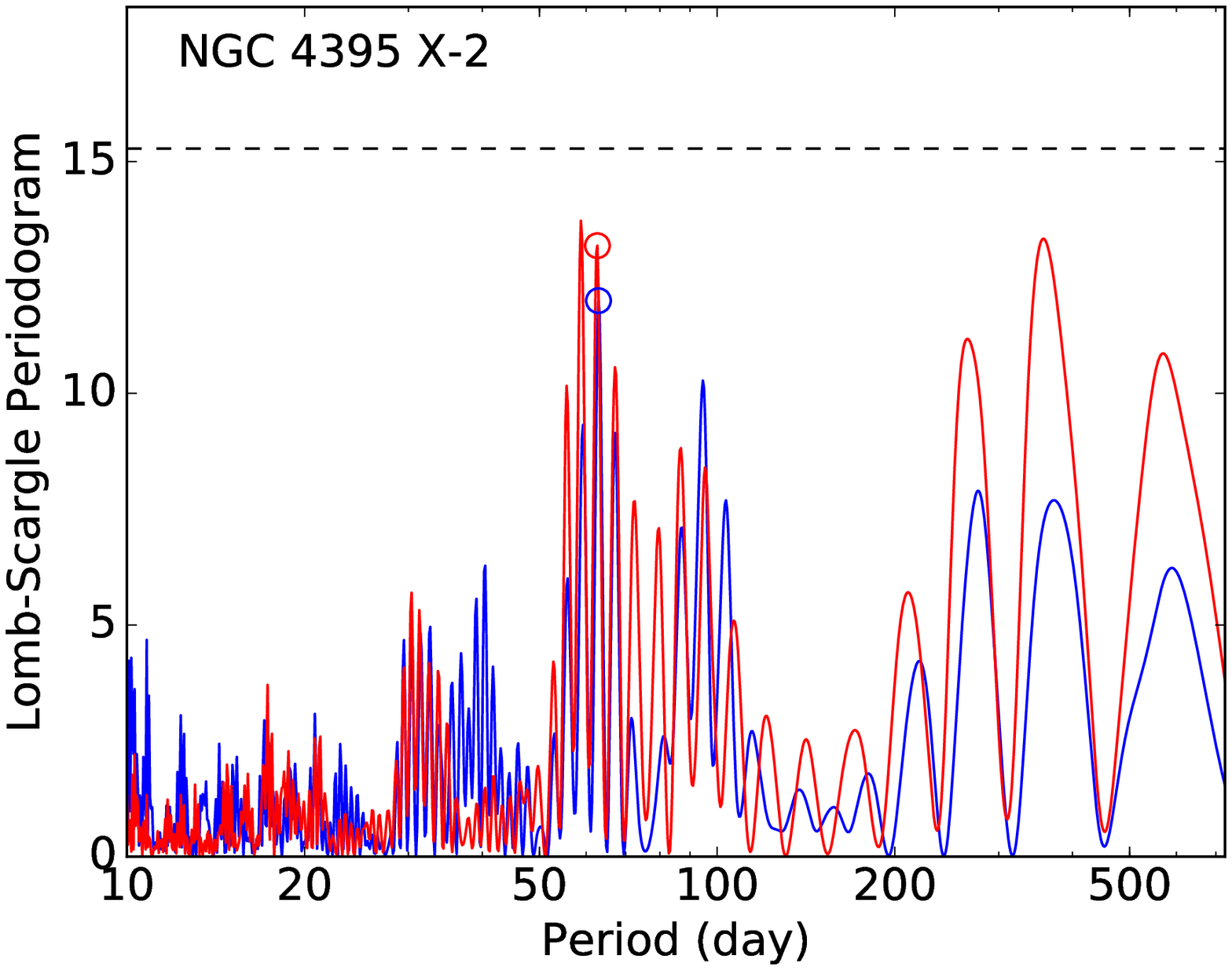}
\includegraphics[width=0.3\textwidth]{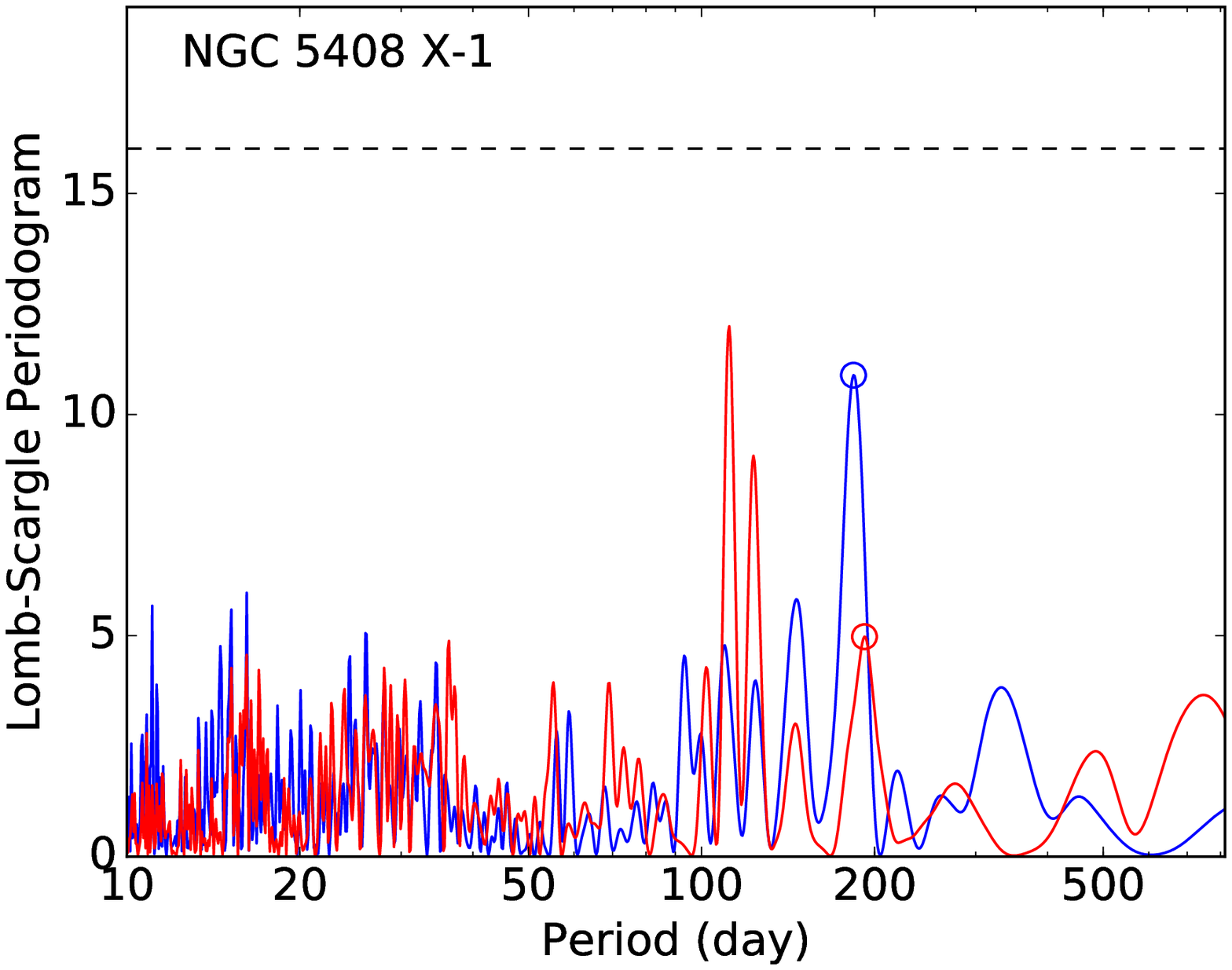}\\
\includegraphics[width=0.3\textwidth]{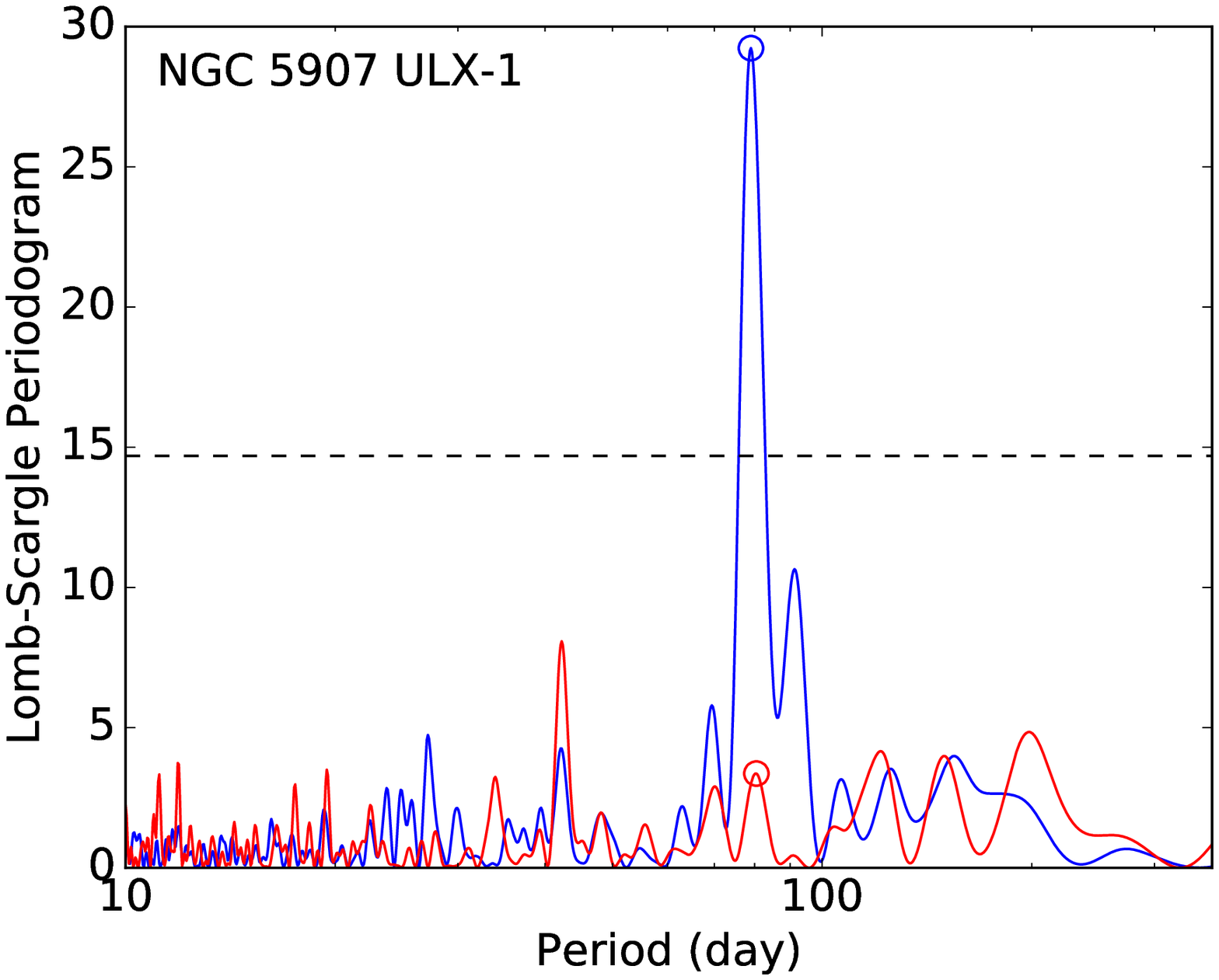}
\includegraphics[width=0.3\textwidth]{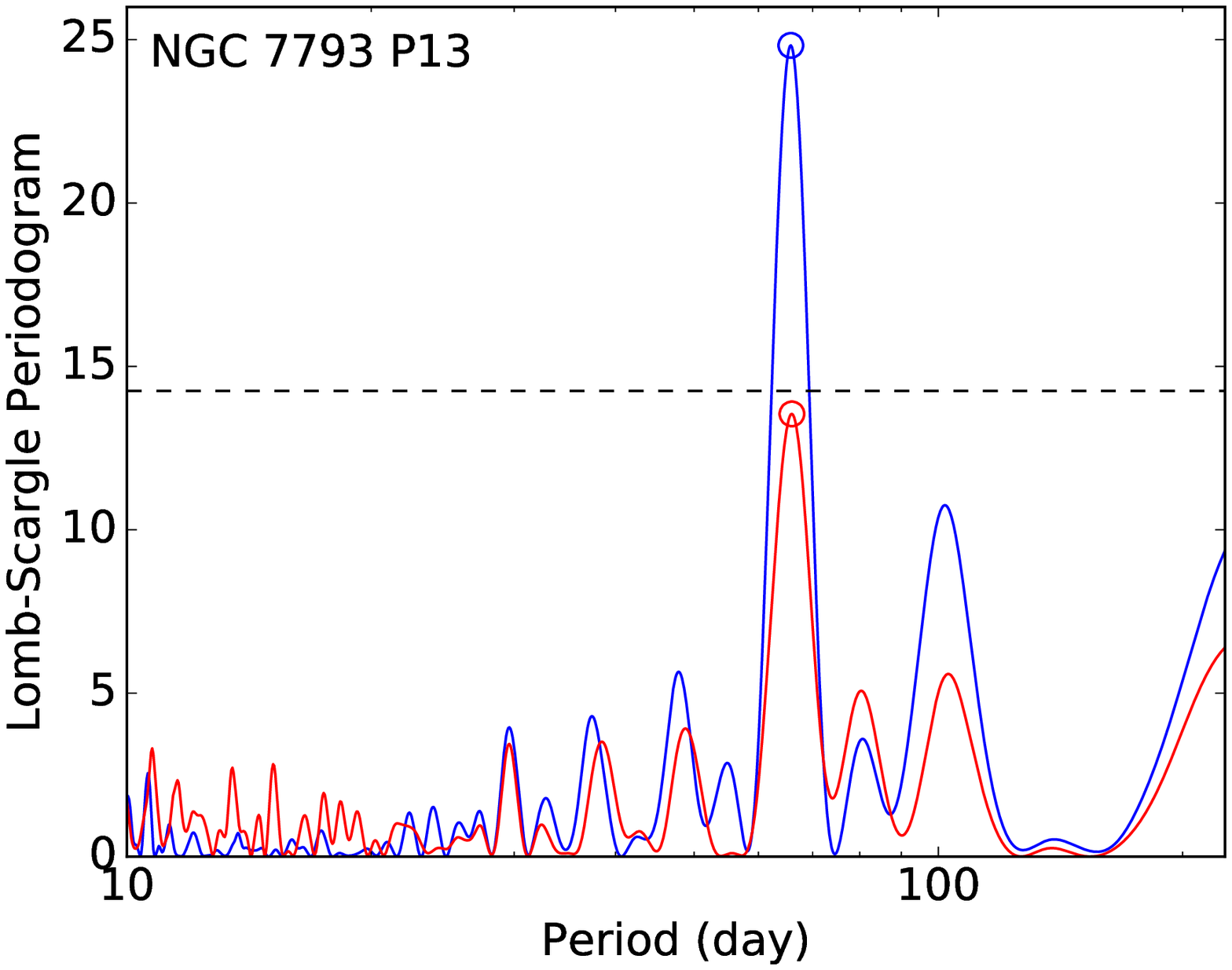}
\includegraphics[width=0.3\textwidth]{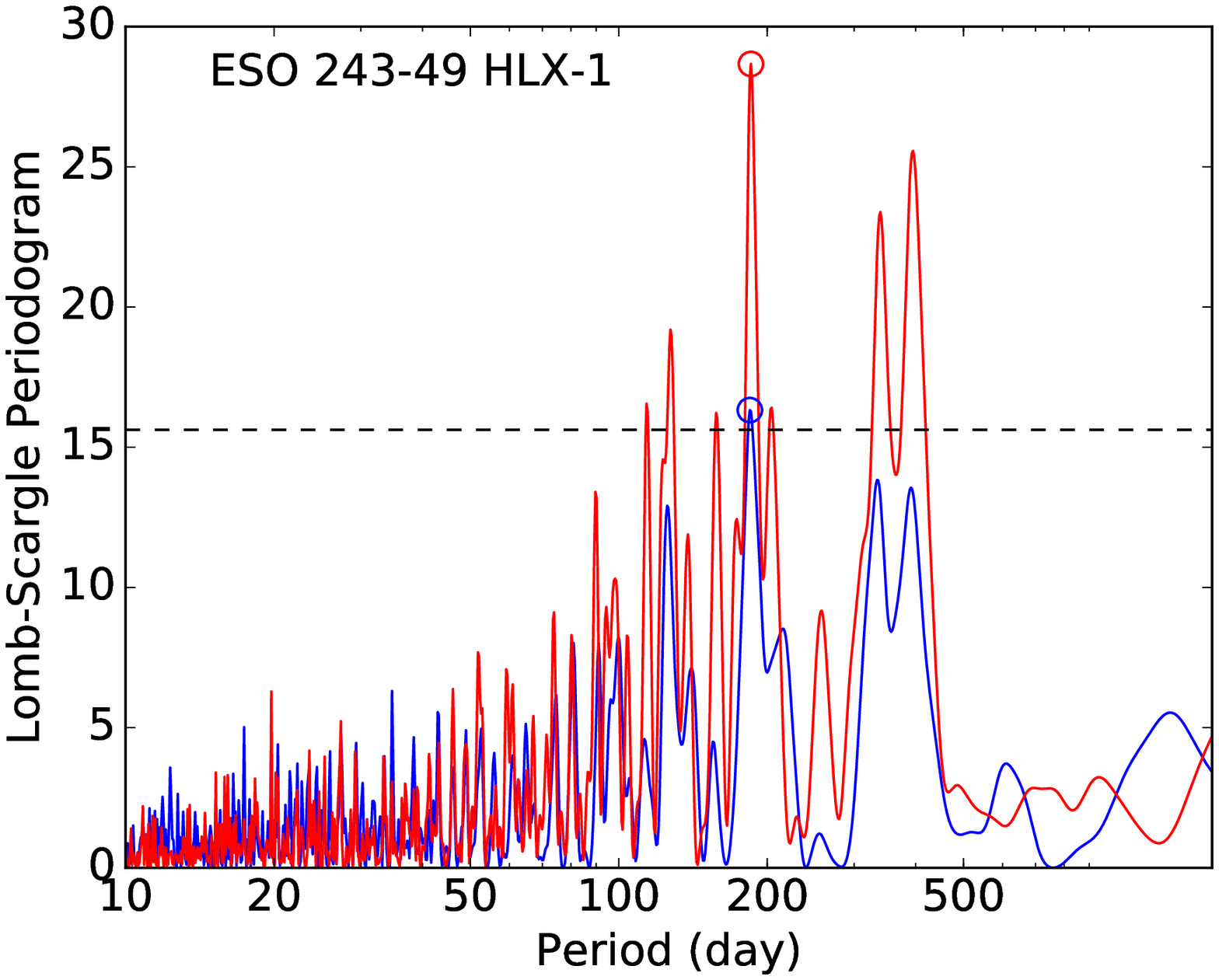}\\
\caption{Lomb--Scargle periodograms for ULXs in the sample with high quality light curves.  The dashed line indicates a 4$\sigma$ detection above the white noise. The blue line is for the hard band (1--10 keV) and the red is for the soft band (0.3--1 keV). The blue circle indicates the peak of the power in the hard band; if there are possible harmonics or multiple significant peaks, the one of the shortest period is chosen. The red circle marks the corresponding peak in the soft band.
\label{fig:pd}}
\end{figure*}

\begin{figure*}
\includegraphics[width=0.3\textwidth]{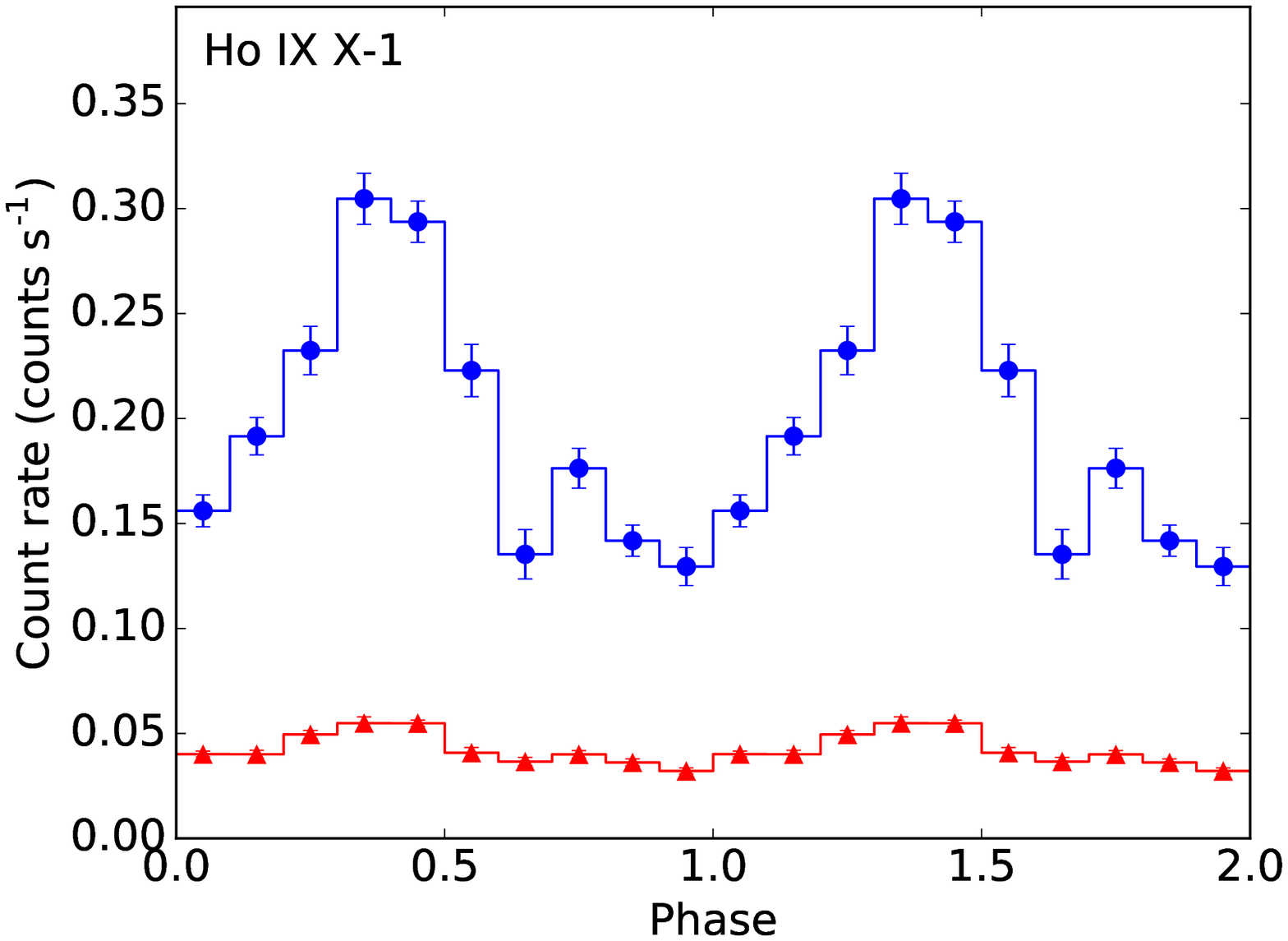}
\includegraphics[width=0.3\textwidth]{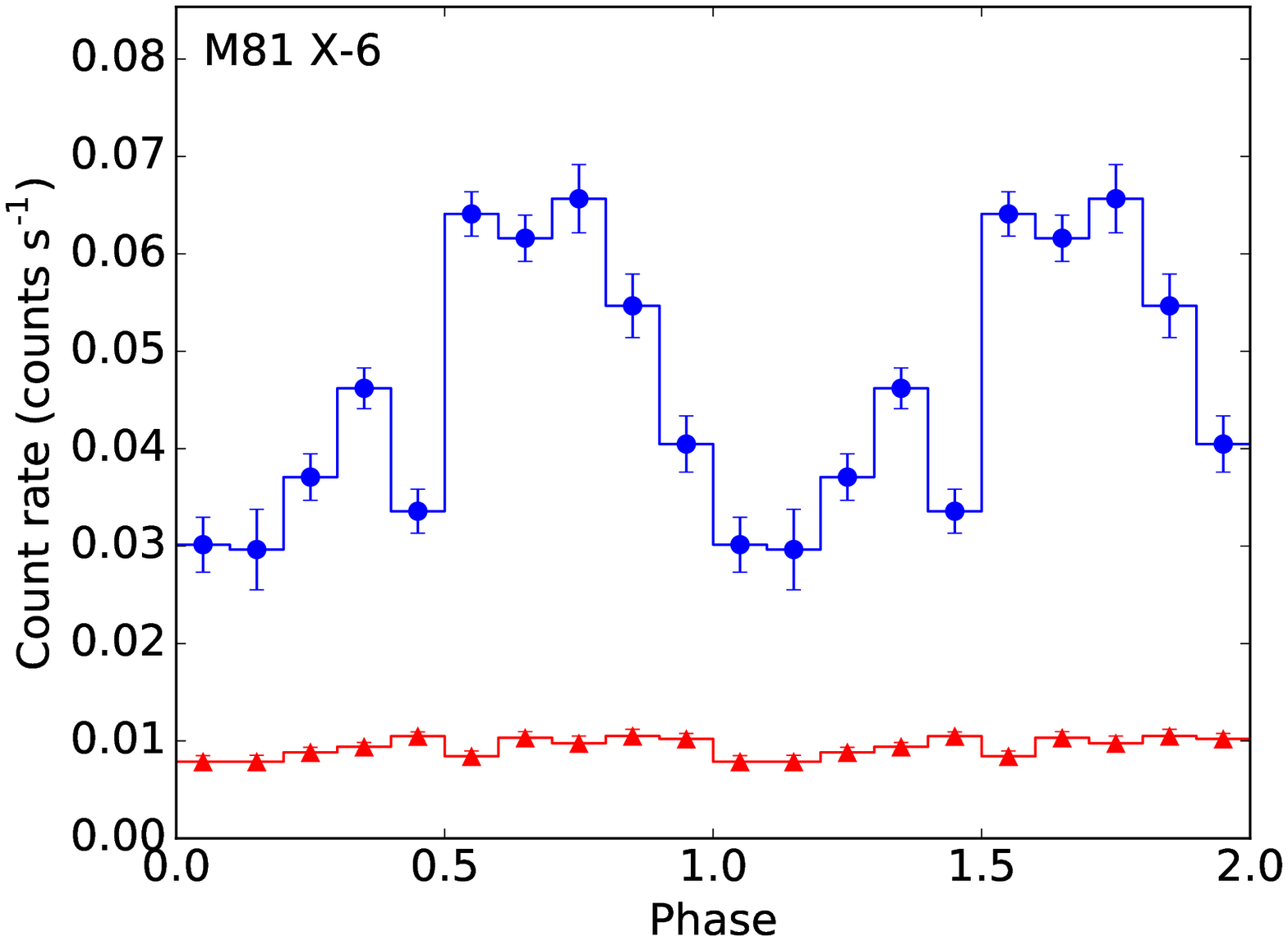}
\includegraphics[width=0.3\textwidth]{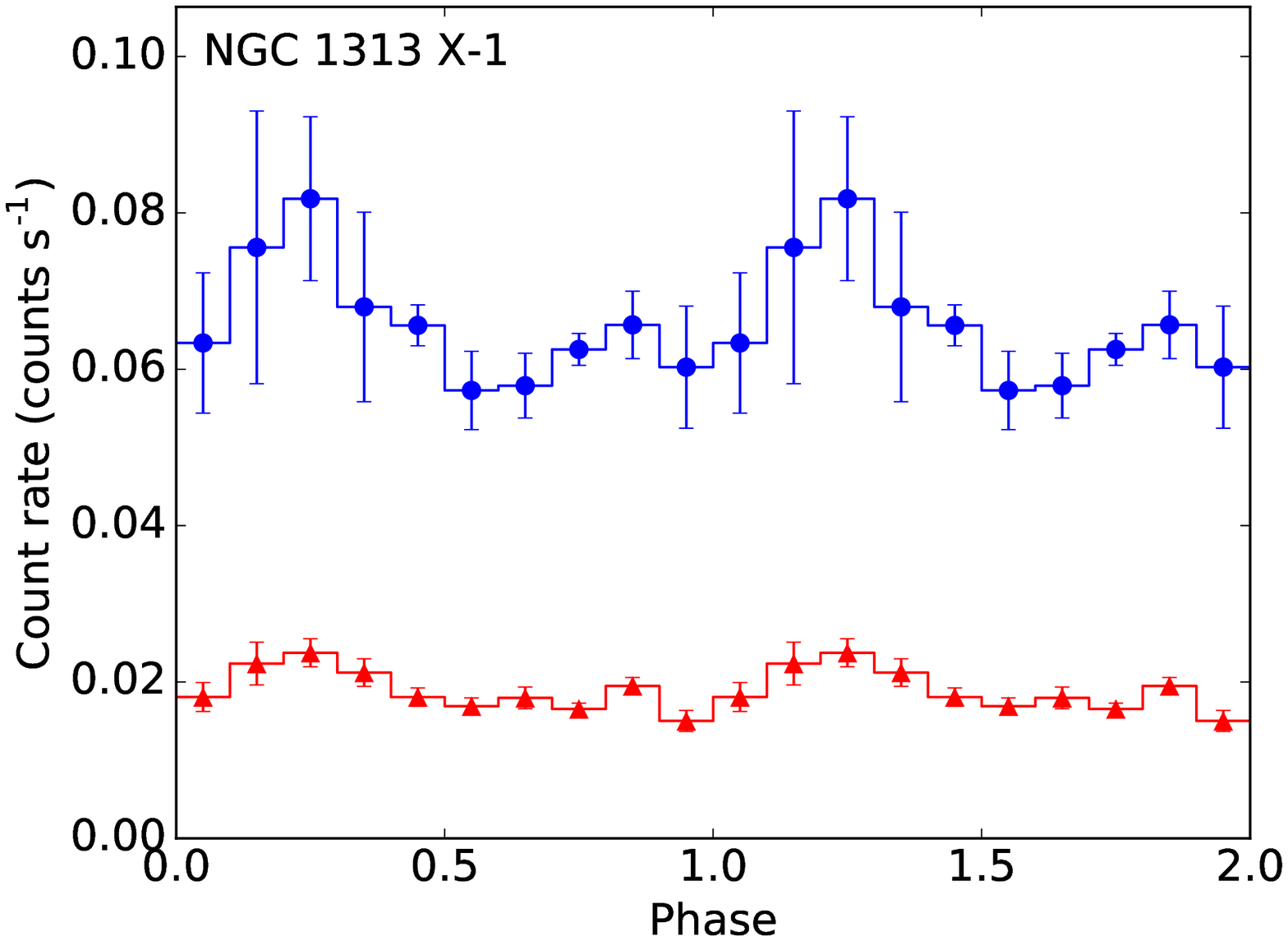}\\
\includegraphics[width=0.3\textwidth]{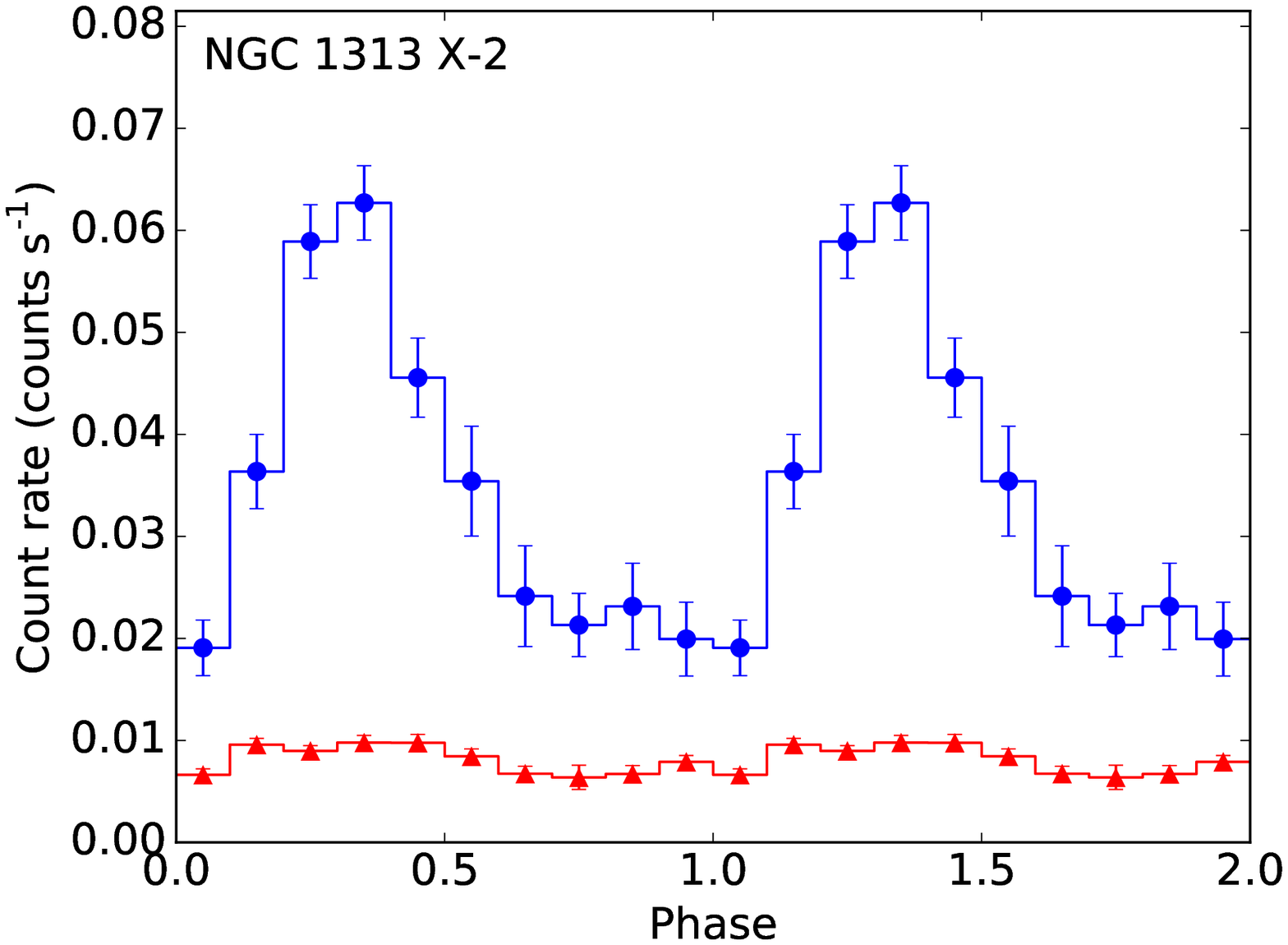}
\includegraphics[width=0.3\textwidth]{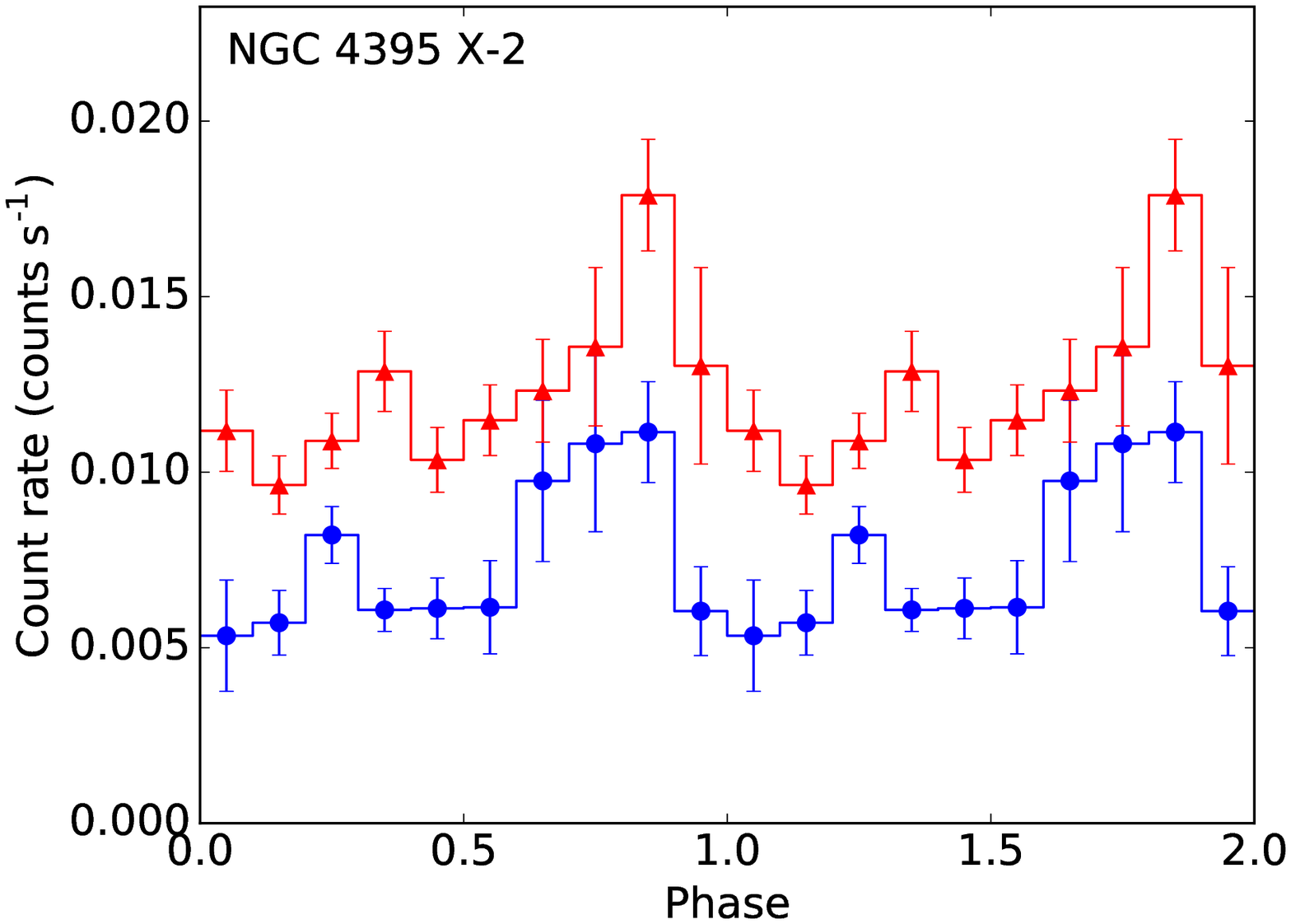}
\includegraphics[width=0.3\textwidth]{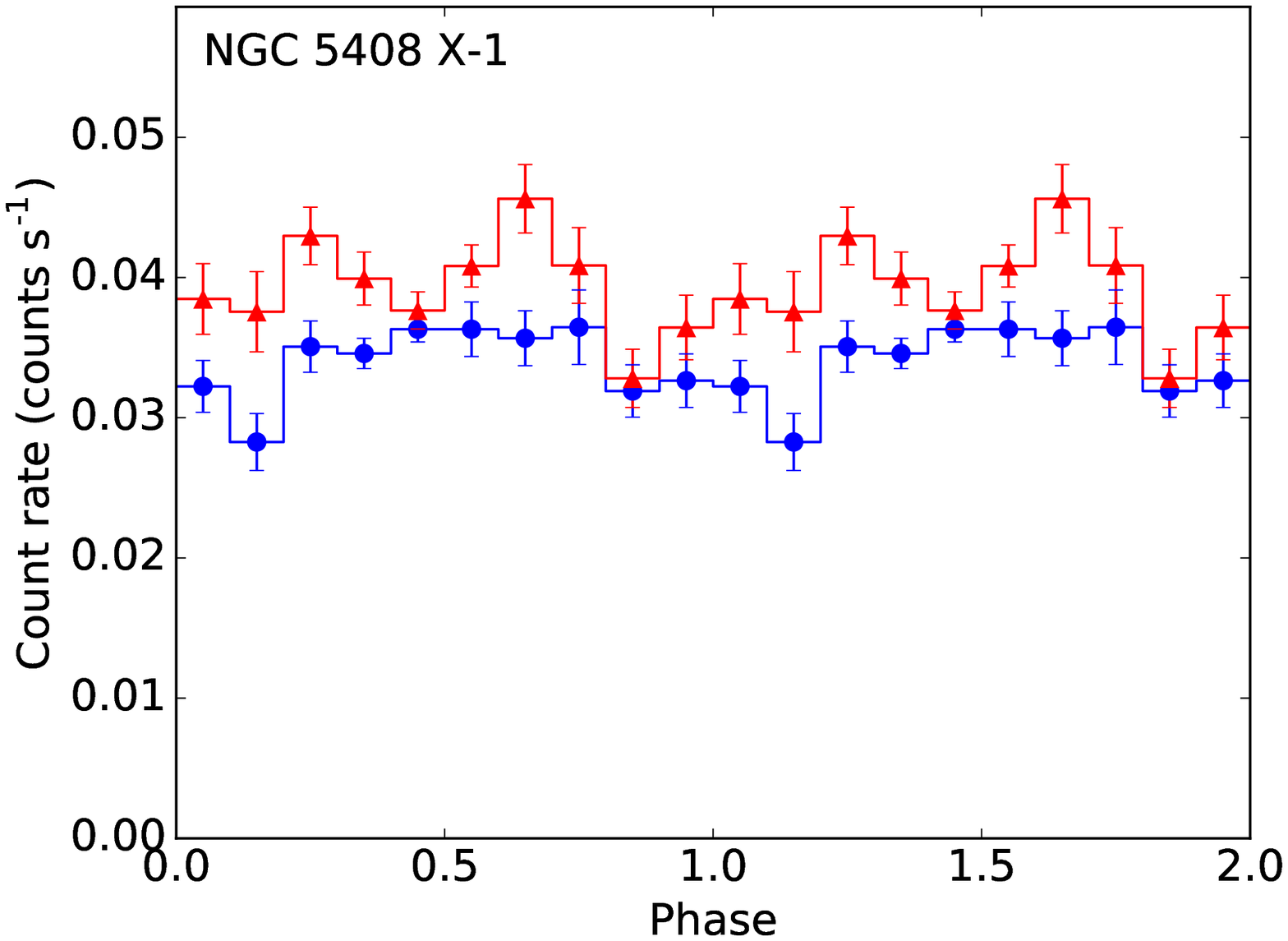}\\
\includegraphics[width=0.3\textwidth]{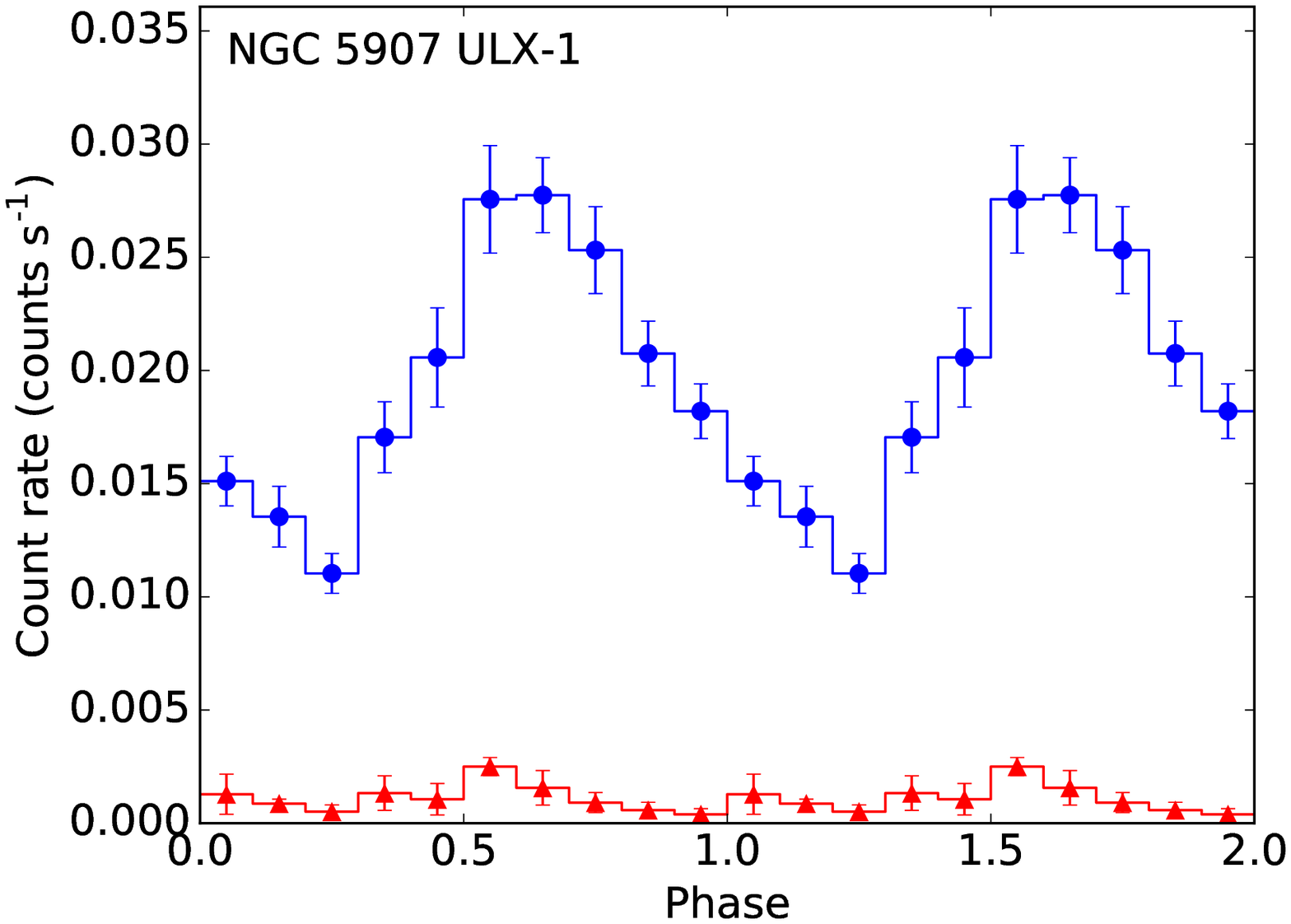}
\includegraphics[width=0.3\textwidth]{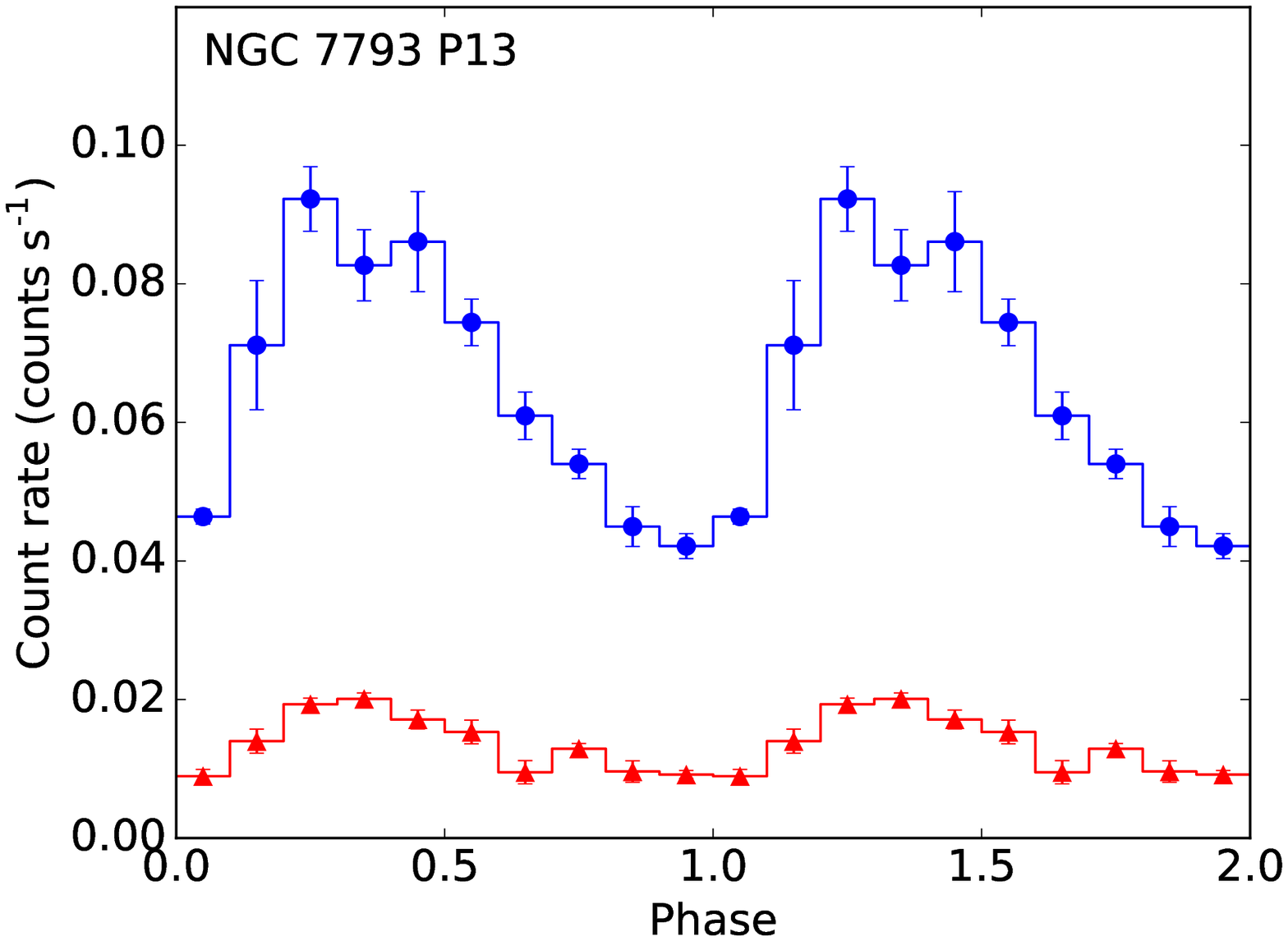}
\includegraphics[width=0.3\textwidth]{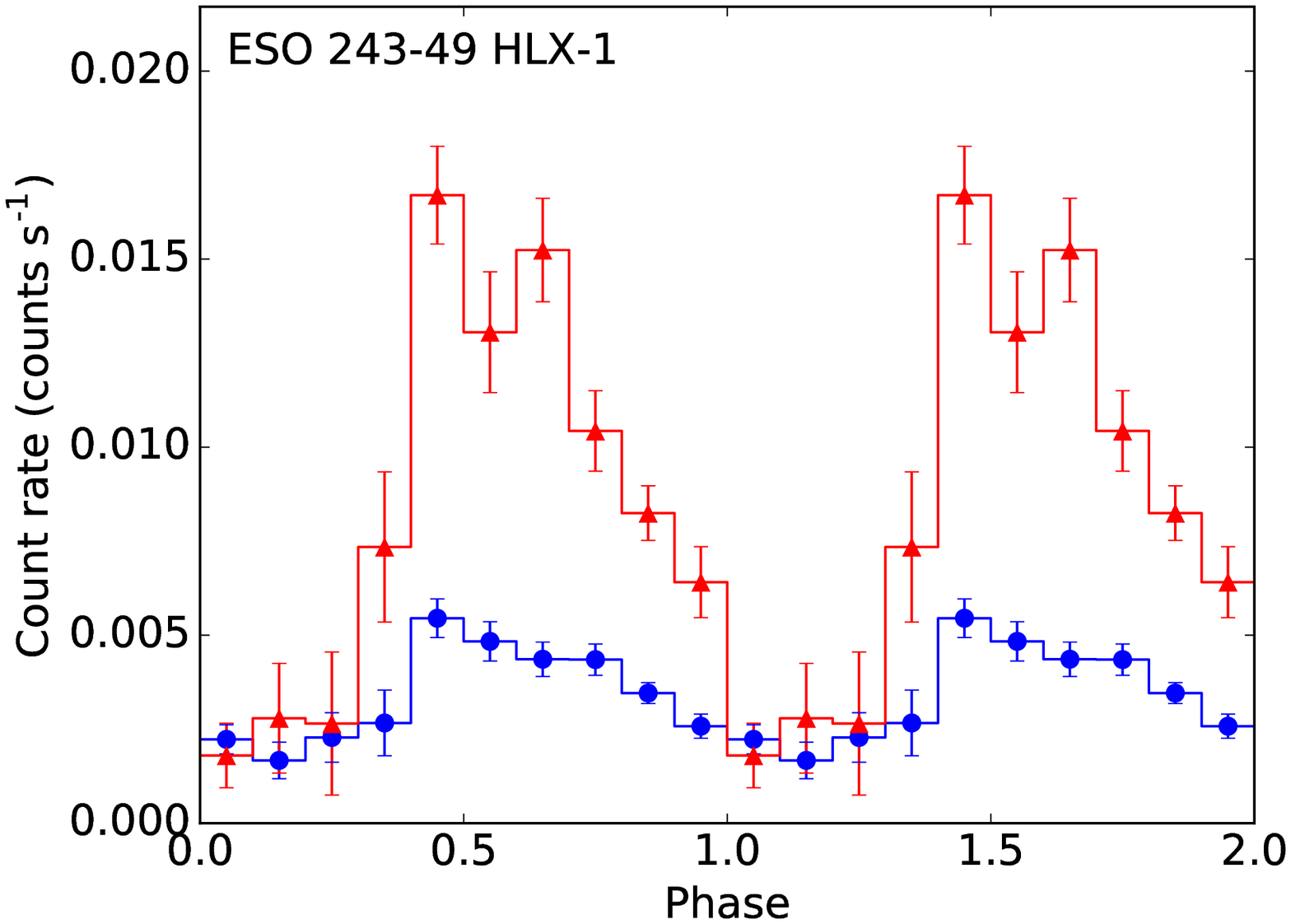}\\
\caption{Light curves folded at their most likely period (indicated by the blue circle in Figure~\ref{fig:pd}).  The blue is for the hard band (1--10 keV) and the red is for the soft band (0.3--1 keV).
\label{fig:fold}}

\end{figure*}

Some of the findings in this paper have been reported in the literature. For NGC 5408 X-1, no significant periodicity is found but the two peaks (near 110~days and 190~days) shown in Figure~\ref{fig:pd} are consistent with previous reports \citep{Grise2013,Pasham2013,Lin2015}. \citet{Lin2015} detected a significant quasi-periodic signal around $\sim$625~days for Ho IX X-1; we detected this peak but considered the one at 266~days as the tentative peak, as there are only four cycles to support the longer period. They also found that the periodicity was not stable for M81 X-6, varying from $\sim$370~days in an earlier segment to $\sim$110~days later, which was adopted by this work. For HLX-1, the detected period was actually due to recurrent outbursts \citep{Yan2015}, which have been found evidently delayed in recent cycles, see Figure~\ref{fig:lc}. Therefore, although a strong peak in the periodogram in the soft band is detected, we do not consider it to be a quasi-periodic signal. \citet{Luangtip2016} fitted the XRT spectra of Ho IX X-1 with a power-law model and found that the ULX exhibited a similar bimodal distribution on the power-law photon index versus the flux plane, consistent with our results. They also attribute the degeneracy between the spectral shape and luminosity as due to source precession. Here, we combine both the temporal and spectral behaviors of these sources to confine the nature of ULXs.

\begin{figure*}
\includegraphics[width=0.3\textwidth]{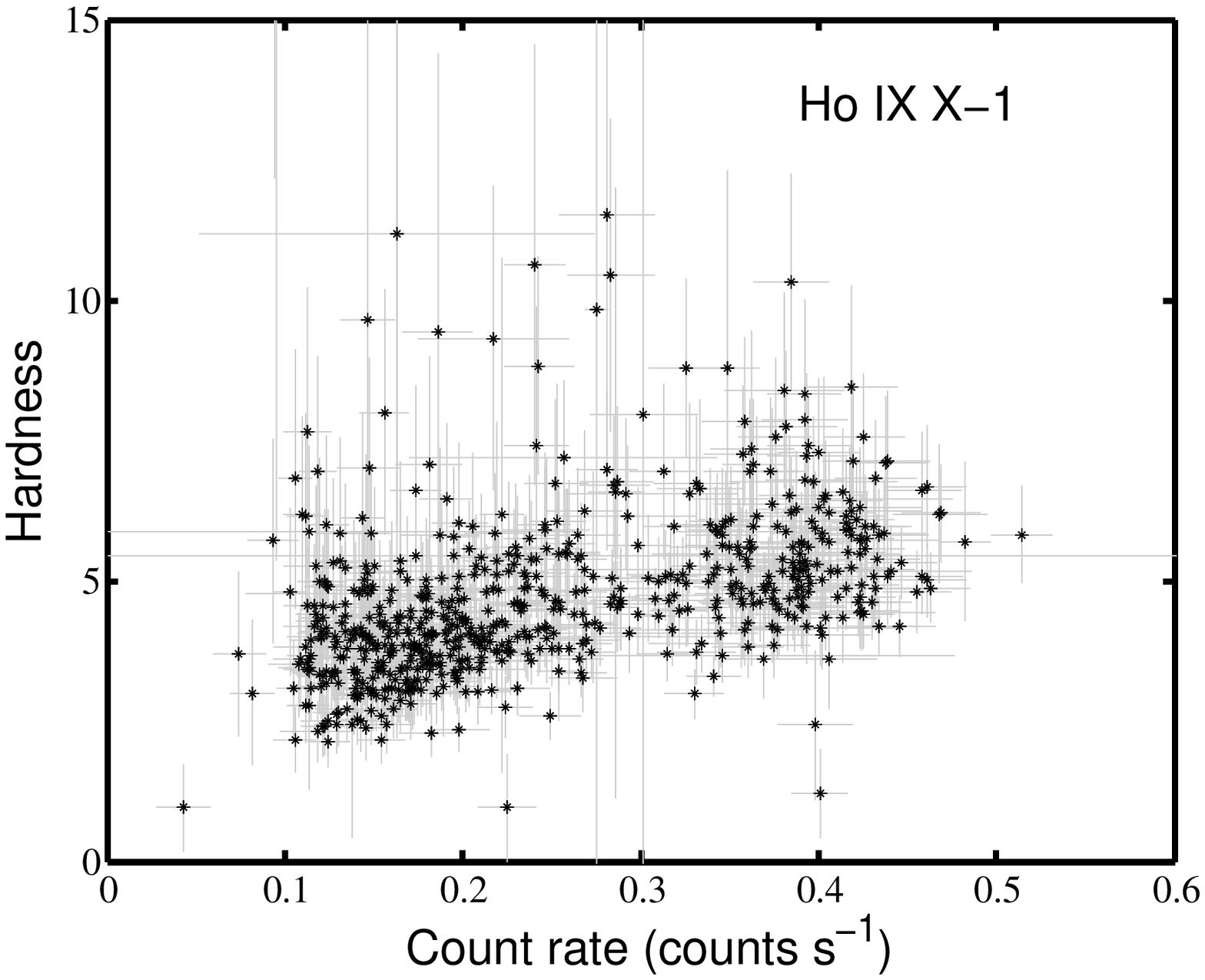}
\includegraphics[width=0.3\textwidth]{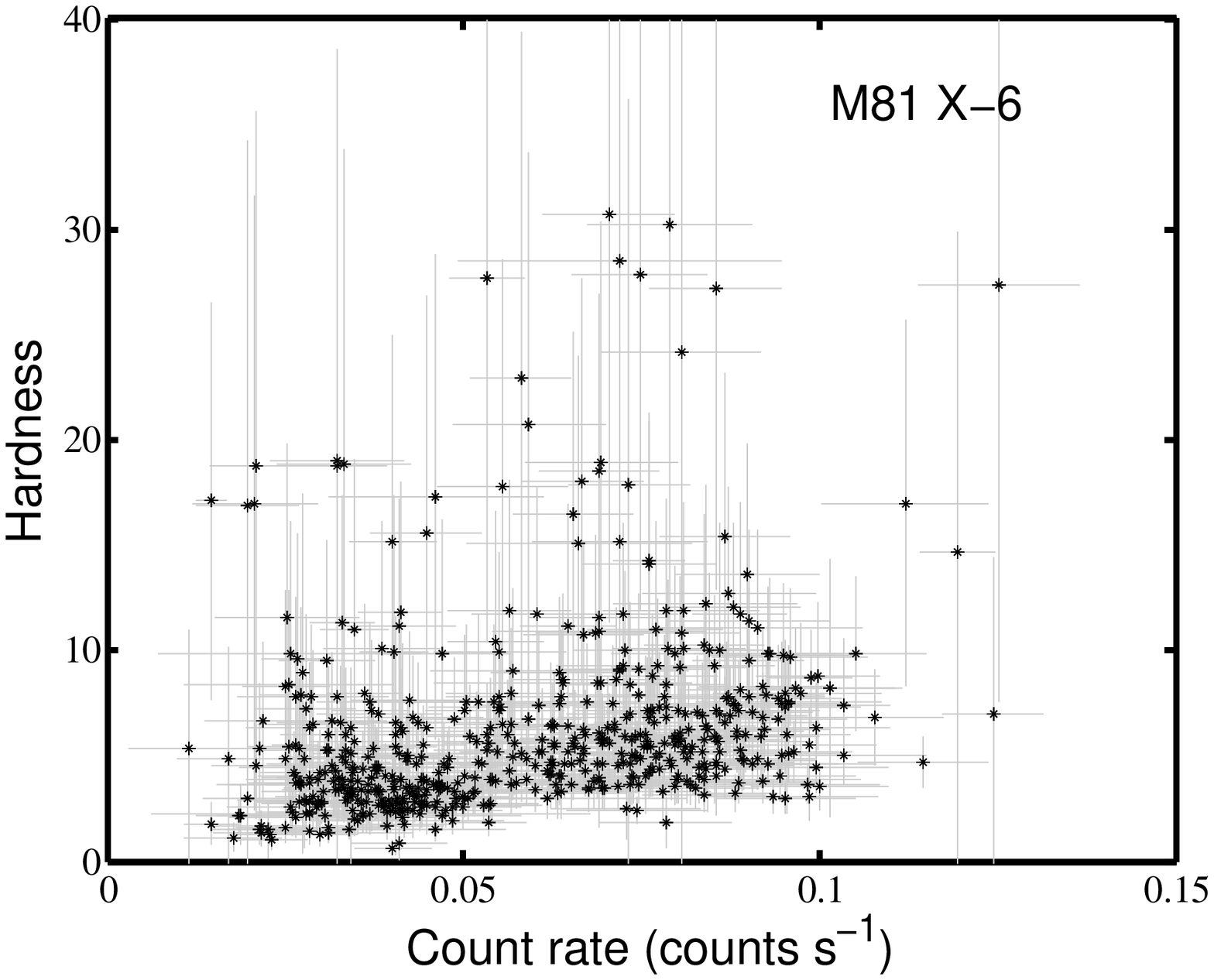}
\includegraphics[width=0.3\textwidth]{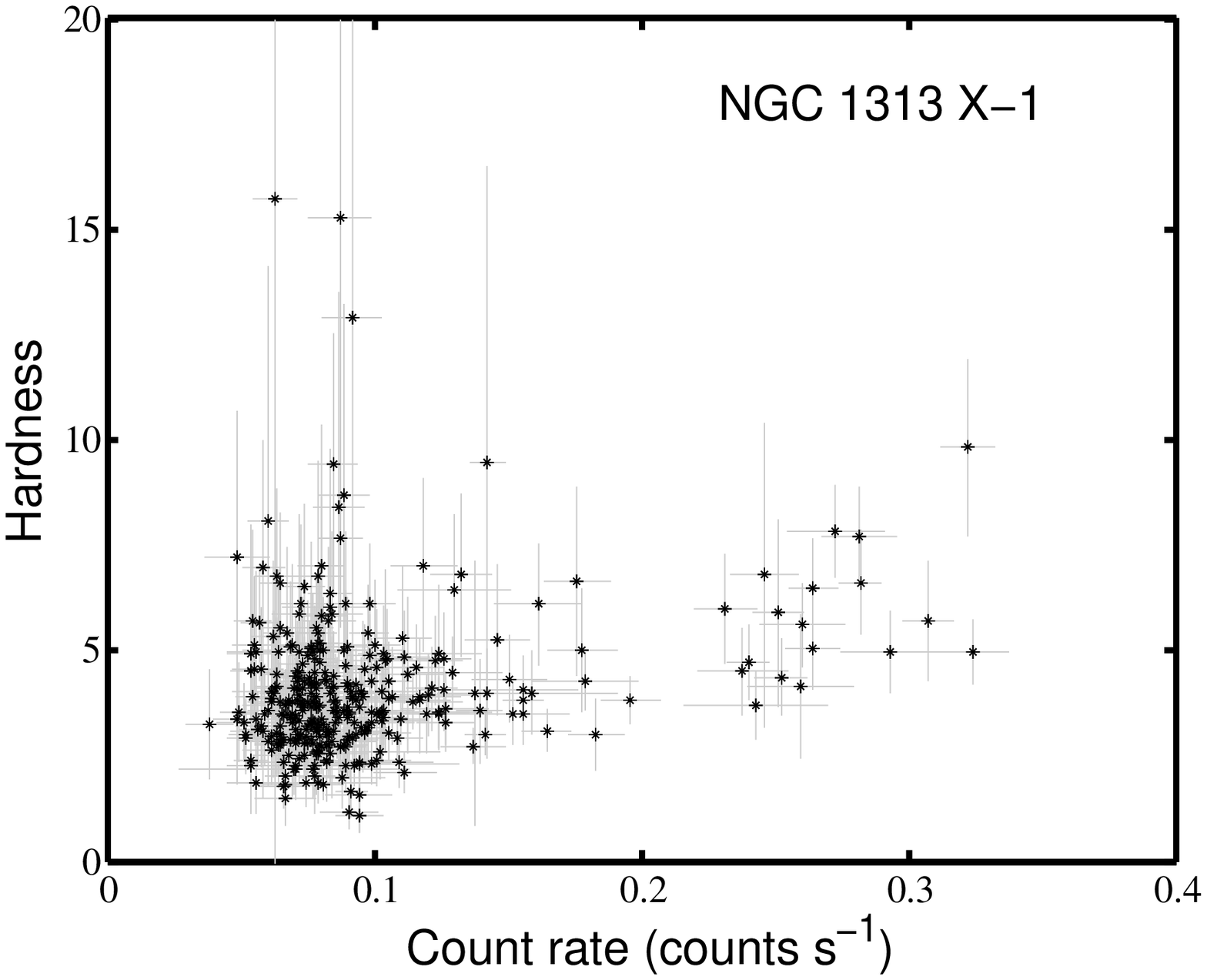}\\
\includegraphics[width=0.3\textwidth]{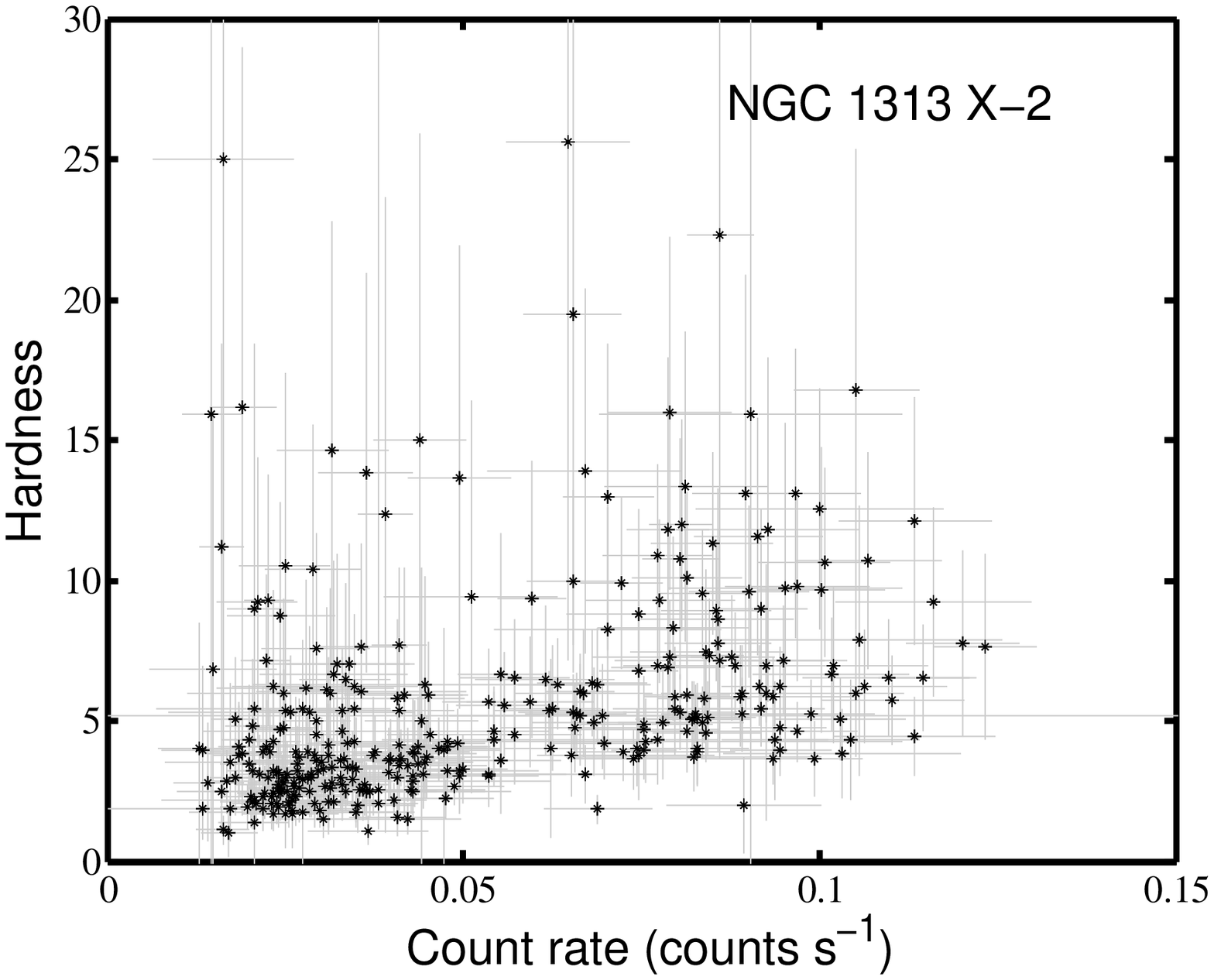}
\includegraphics[width=0.3\textwidth]{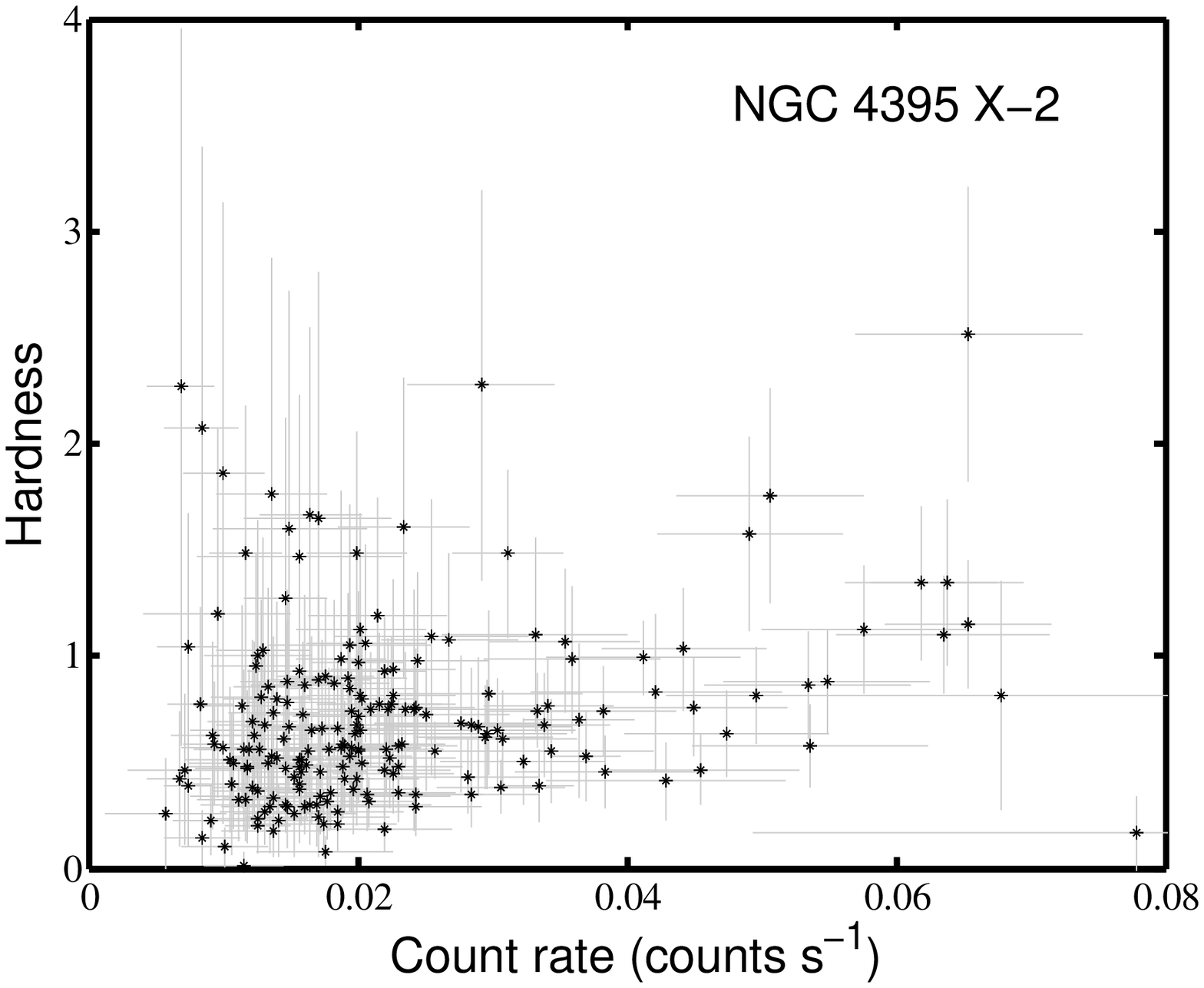}
\includegraphics[width=0.3\textwidth]{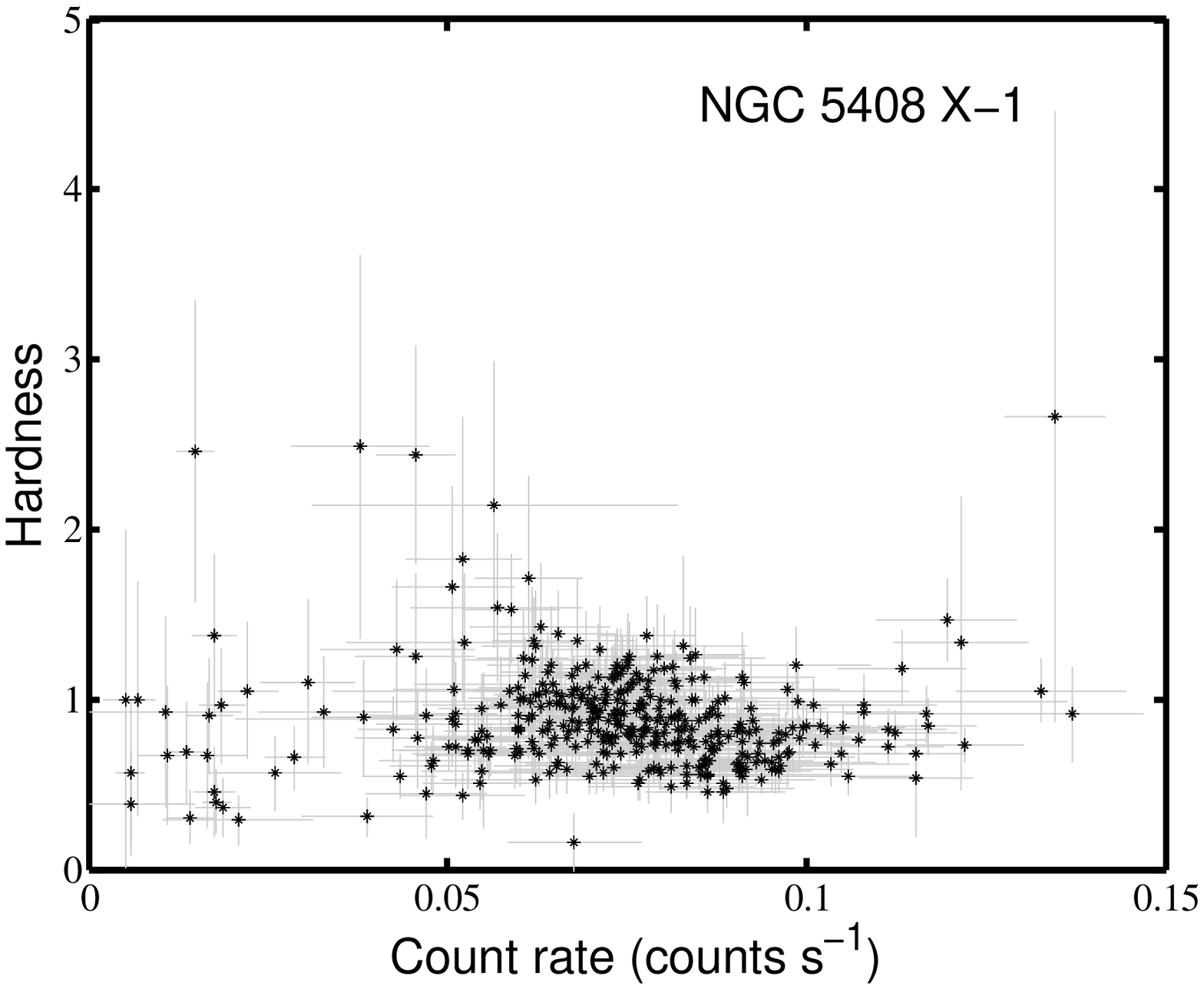}\\
\includegraphics[width=0.3\textwidth]{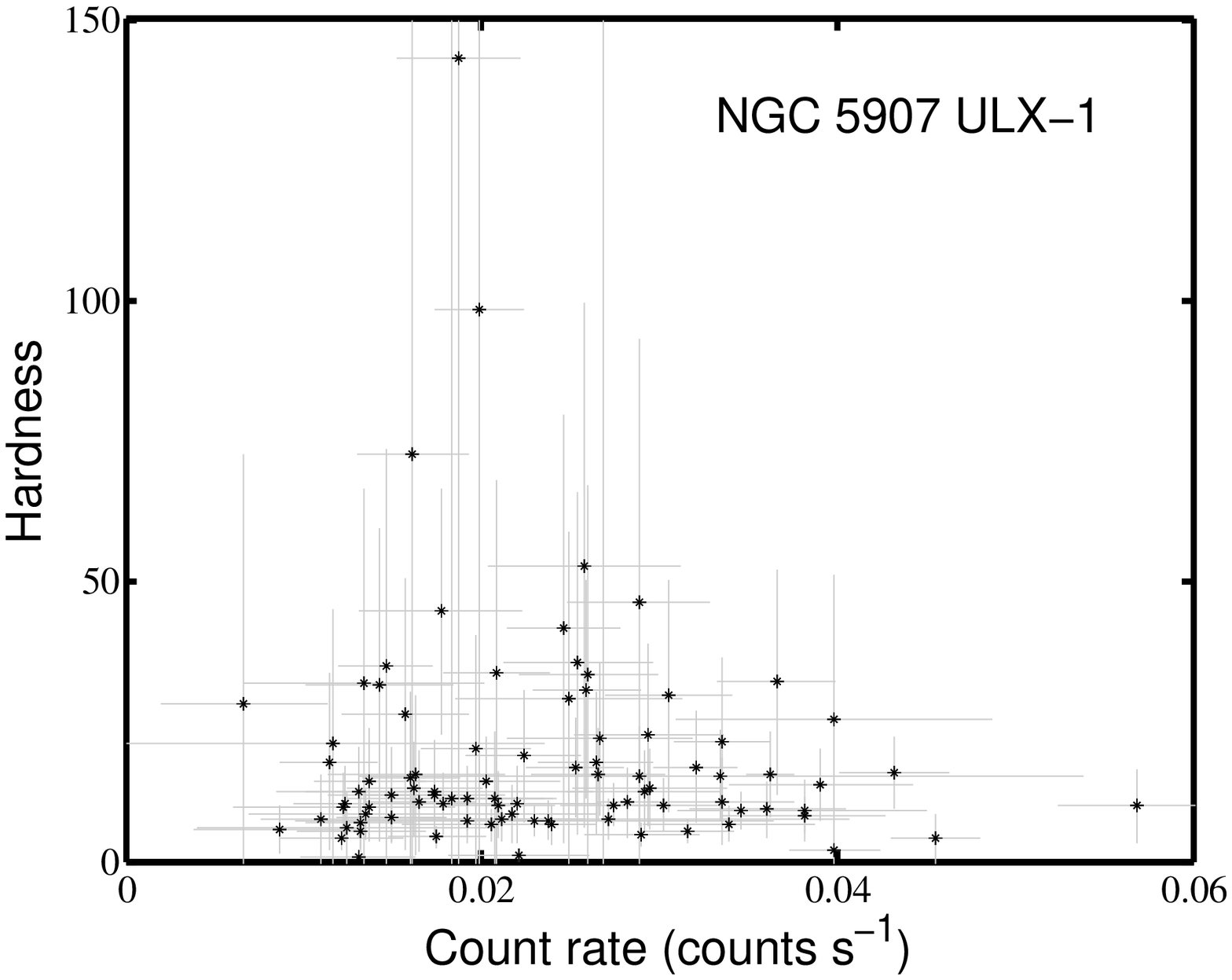}
\includegraphics[width=0.3\textwidth]{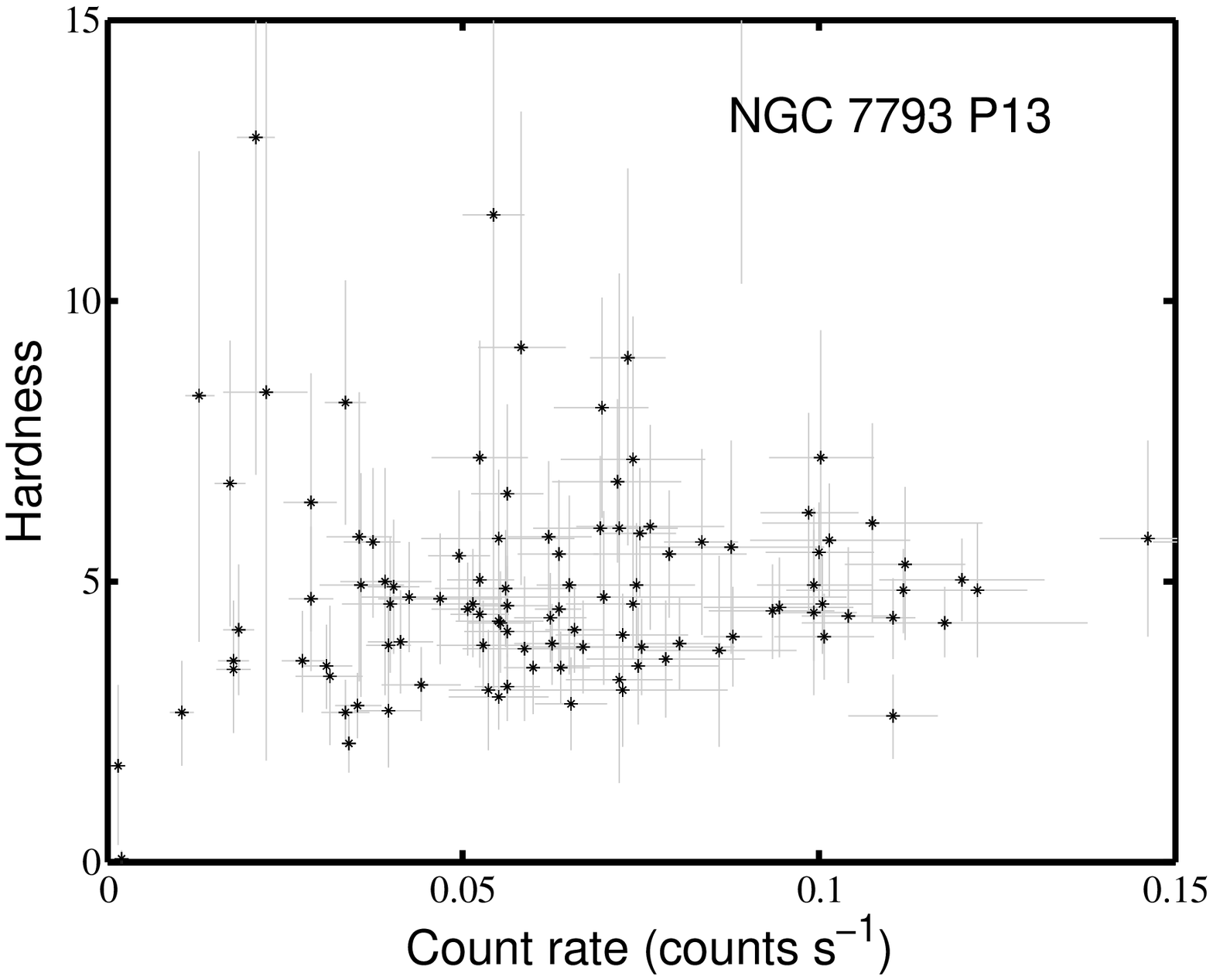}
\includegraphics[width=0.3\textwidth]{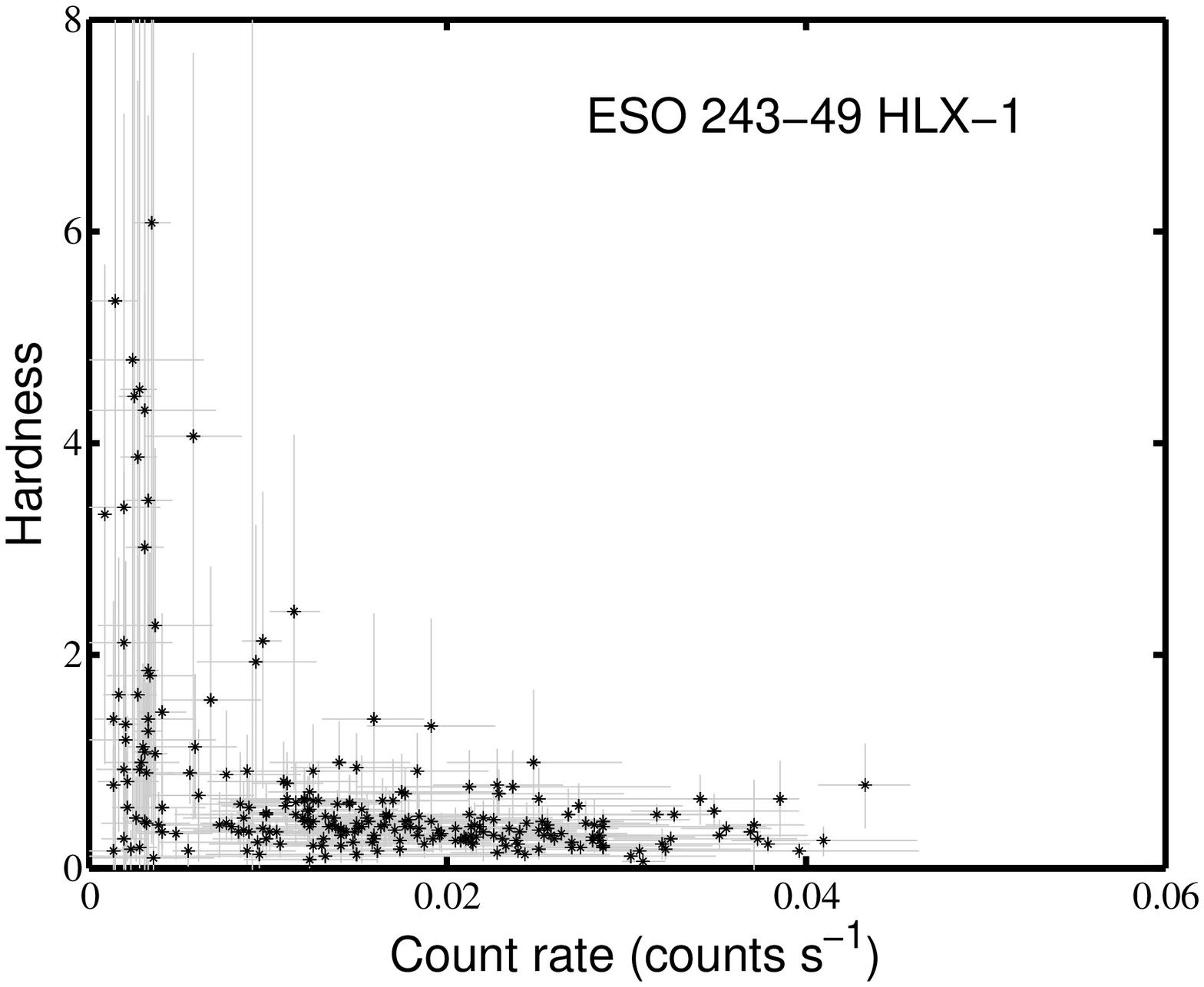}\\
\caption{HIDs for ULXs in the sample. The hardness is defined as the ratio of XRT count rates between the hard band (1--10 keV) and the soft band (0.3--1 keV). The intensity refers to the 0.3--10 keV count rate measured with XRT.  Each point represents a {\it Swift} observation and the errors are of 1$\sigma$.
\label{fig:hid}}
\end{figure*}

The results clearly classify the ULXs into two categories. Four ULXs (Ho IX X-1, M81 X-6, NGC 1313 X-1, and X-2) exhibit a strong (or moderate for NGC 1313 X-1) modulation in the hard band, however, at the same period, the modulation in the soft band is much weaker or absent. The sources showing such a feature also display a bimodal distribution on the HID; a similar bimodal feature can also be reflected on the flux distribution. Two ULXs (NGC 4395 X-2 and NGC 5408 X-1) do not show a significant modulation, nor do they have a bimodal distribution on the HID.  These two sources are the softest among all, see the HID in Figure~\ref{fig:hid} (we note that the they are indeed the softest after correction for absorption).  For such soft ULXs, whether or not a cut at 1 keV can separate the two emission components is a question; a cut at higher energies will lead to less photons in the hard band and difficulties in detection.

We argue that these results can be well explained as due to supercritical accretion with precession. As mentioned in the Introduction, most ULXs are thought to be systems undergoing supercritical accretion. A supercritical accretion flow around a stellar mass black hole may produce two emitting components, a soft, quasi-blackbody component with a temperature of several tenths of a keV arising from the photosphere of the outflow that covers the outer disk, and a hard, Comptonized component from the inner accretion disk \citep[below the inner photospheric radius, see][]{Poutanen2007}. Such a picture has been generally confirmed by recent numerical simulations \citep{Ohsuga2011,Jiang2014,Scadowski2016}, and can be used to explain the spectral-timing behaviors of bright ULXs \citep{Middleton2015}. As the emission from the inner disk can only be seen via a central funnel enclosed by the thick outflow, the hard X-rays could be collimated, while the soft X-rays are almost isotropic. In this picture, precession of the accretion flow will cause a larger modulation factor on the hard X-ray flux than on the soft X-ray flux. Whether or not there is a bimodal distribution of the hard X-ray flux depends on its angular distribution. For instance, in an extreme condition where the hard X-rays are strictly collimated by the wall of the central cone, a bimodal distribution is naturally seen when the hard X-ray beam enters and leaves the line of sight.  In practice, the hard X-ray emission may have an angular distribution much wider than the open angle of the funnel, due to multiple scattering below the scattering photosphere \citep{Jiang2014}.  Thus, the folded light curve may be useful to model the hard X-ray angular distribution, or the geometry of the accretion flow.  Upon the assumptions above, our results suggest that the hard X-ray emission is more or less beamed and very likely originated from the central funnel.

The two pulsars NGC 5907 ULX-1 and NGC 7793 P13, like other ULXs, display strong modulation in the hard X-ray band but not in the soft band. However, it does not show a bimodal distribution on the HID. As discussed above, the bimodal distribution depends on how the hard X-ray emission is modulated. The folded light curve of the pulsar displays a smooth transition from the peak to valley like a triangle waveform. This may imply that the pulsar has a different shape or mechanism of beaming, indicating that other ULXs may contain a black hole instead of a neutron star. We want to emphasize that this preliminary conclusion is based on only two sources, and the number of observations for these two pulsars are small compared with other ULXs. So, along with the fact that ESO 243-49 HLX-1 does not show a bimodal feature, the conclusion is tentative and should be further tested with more observations.


The hyperluminous X-ray source ESO 243-49 HLX-1 has been argued to be a good candidate for an IMBH \citep{Farrell2009}. It shows possible periodicity, stronger in the soft band than in the hard band, that is obviously due to recurrent outbursts \citep{Yan2015}, which have evidently been found to have been delayed in recent cycles. This clearly distinguishes this source from other ULXs.

To conclude, a hard X-ray quasi-periodicity along with a nearly constant soft X-ray flux is quite consistent with the scenario described above that the sources are undergoing supercritical accretion with precession. The absence of such features seems to belong to the softest ULXs, which are argued to be more inclined systems \citep{Sutton2013,Middleton2015}, where the precession angle may be smaller than the inclination angle, even if the precession occurs. The results suggest that precession may be ubiquitous in supercritical accretion systems like SS~433, as a natural consequence of strong radiation pressure and outflows.

\acknowledgments We thank the anonymous referee for useful comments, and Junxian Wang and Zhenyi Cai for helpful discussions. This work is supported by the National Natural Science Foundation of China under grants 11673013, 11573023, and 11633003, and the National Program on Key Research and Development Project (Grant No.\ 2016YFA040080X).


\clearpage

\end{document}